\documentclass[fleqn,usenatbib]{mnras}

\newcommand{\ebv}{E($B$-$V$)}

\newcommand{\halpha}{H$\alpha$}

\newcommand{\gaia}{\emph{Gaia}}
\newcommand{\ktwo}{\emph{K2}}

\newcommand{\kepler}{\emph{Kepler}}

\newcommand{\wise}{\emph{WISE}}
\newcommand{\spitzer}{\emph{Spitzer}}

\newcommand{\ks}{\mbox{km\,s$^{-1}~$}}

\newcommand{\msun}{M$_{\odot}~$}
\newcommand{\msune}{M$_{\odot}$}
\newcommand{\rsun}{R$_{\odot}~$}
\newcommand{\lsun}{L$_{\odot}~$}

\newcommand{\mjup}{M$_{\rm JUP}~$}

\newcommand{\teff}{\ensuremath{T_{\rm eff}}}

\newcommand{\logg}{\ensuremath{\log{g}}}

\newcommand{\lbol}{$L_*$}

\newcommand{\comover}{2M043726}

\usepackage[T1]{fontenc}
\usepackage{ae,aecompl}
\usepackage{tablefootnote}

\usepackage{graphicx}	
\usepackage{amsmath}	
\usepackage{amssymb}	

\usepackage{newtxtext,newtxmath}
\usepackage{lineno,multirow}

\title[]{Zodiacal Exoplanets in Time (ZEIT) XII: A Directly-Imaged Planetary-Mass Companion to a Young Taurus M Dwarf Star}
\author[Gaidos et al.]{E. Gaidos$^{1,2,3}$
\thanks{E-mail: gaidos@hawaii.edu.}
\thanks{Visiting Astronomer at the Infrared Telescope Facility, which is operated by the University of Hawaii under contract 80HQTR19D0030 with the National Aeronautics and Space Administration.}, 
T. Hirano$^{4,5}$, 
A. L. Kraus$^{6}$, 
M. Kuzuhara$^{4,5}$,
Z. Zhang$^{6}$,
R. A. Lee$^{1}$,
\newauthor
M. Salama$^{7}$,
T. A. Berger$^{8}$, 
S. K. Grunblatt$^{8,9,10}$,
M. Ansdell$^{11}$, 
M. C. Liu$^{8}$,
H. Harakawa$^{12}$,
\newauthor
K. W. Hodapp$^{7}$, 
S. Jacobson$^{7}$,
M. Konishi$^{13}$, 
T. Kotani$^{4,5,14}$,
T. Kudo$^{12}$, 
T. Kurokawa$^{4,15}$,
\newauthor
J. Nishikawa$^{5,14,4}$,
M. Omiya$^{4,5}$,
T. Serizawa$^{15}$,
M. Tamura$^{4,5,16}$, 
A. Ueda$^{5}$,
\& S. Vievard$^{12}$\\
$^{1}$Department of Earth Sciences, University of Hawai'i at M\={a}noa, 1680 East-West Rd, Honolulu, HI  96822, USA\\
$^{2}$Center for Space and Habitability, University of Bern, Gesellschaftsstrasse 6, 3012 Bern, Switzerland\\
$^{3}$Institute for Astrophysics, University of Vienna, T\"{u}rkenschanzstrasse 17, 1180 Vienna, Austria\\
$^{4}$Astrobiology Center, 2-21-1 Osawa, Mitaka, Tokyo 181-8588, Japan\\
$^{5}$National Astronomical Observatory of Japan, NINS, 2-21-1 Osawa, Mitaka, Tokyo 181-8588, Japan\\
$^{6}$Department of Astronomy, University of Texas at Austin, 2515 Speedway, Austin, TX 78712, USA\\
$^{7}$Institute for Astronomy, University of Hawai'i at M\={a}noa, 2680 Woodlawn Dr, Honolulu, HI 96822, USA\\
$^{8}$Institute for Astronomy, University of Hawai'i at Hilo, 640 N. Aohoku Place, Hilo, HI 96720, USA\\
$^{9}$American Museum of Natural History, 200 Central Park W, New York, NY 10024, USA\\
$^{10}$Center for Computational Astrophysics, Flatiron Institute, 162 5th Ave, New York, NY 10010, USA\\
$^{11}$NASA Headquarters, 300 E Street, N.W., Washington, DC 20546, USA\\
$^{12}$Subaru Telescope, 650 N. Aohoku Place, Hilo, HI 96720, USA\\
$^{13}$Faculty of Science and Technology, Oita University, 700 Dannoharu, Oita 870-1192, Japan\\
$^{14}$Department of Astronomy, School of Science, The Graduate University for Advanced Studies (SOKENDAI), 2-21-1 Osawa, Mitaka, Tokyo, Japan\\
$^{15}$Tokyo University of Agriculture and Technology, 2-24-16, Naka-cho, Koganei, Tokyo, 184-8588, Japan\\
$^{16}$Department of Astronomy, Graduate School of Science, The University of Tokyo, 7-3-1 Hongo, Bunkyo-ku, Tokyo 113-0033, Japan\\
}

\date{Accepted to 2021 October 14. Received 2021 October 12; in original form 2021 July 30}

\pubyear{2021}

\hypersetup{draft}
\begin{document}
\label{firstpage}
\pagerange{\pageref{firstpage}--\pageref{lastpage}}
\maketitle

\begin{abstract}
We report the discovery of a resolved (0\arcsec.9) substellar companion to a member of the 1-5 Myr Taurus star-forming region. The host star (2M0437) is a single mid-M type (\teff$\approx$3100\,K) dwarf with a position, space motion, and color-magnitude that support Taurus membership, and possible affiliation with a $\sim$2.5 Myr-old sub-group.  A comparison with stellar models suggests a 2-5 Myr age and a mass of 0.15-0.18\msune.  Although \ktwo\ detected quasi-periodic dimming from close-in circumstellar dust, the star lacks detectable excess infrared emission from a circumstellar disk and its \halpha\ emission is not commensurate with accretion.  Astrometry based on three years of AO imaging shows that the companion (2M0437b) is co-moving, while photometry of two other sources at larger separation indicates they are likely heavily-reddened background stars.  A comparison of the luminosity of 2M0437b with models suggests a mass of 3-5 \mjup, well below the deuterium burning limit, and an effective temperature of 1400-1500\,K, characteristic of a late L spectral type.  The $H$-$K$ color is redder than the typical L dwarf, but comparable to other directly detected young planets, e.g. those around HR 8799.  The discovery of a super-Jupiter around a very young, very low mass star challenges models of planet formation by either core accretion (which requires time) or disk instability (which requires mass).   We also detected a second, co-moving, widely-separated (75\arcsec) object that appears to be a heavily-extincted star.  This is certainly a fellow member of this Taurus sub-group and statistically likely to be a bound companion.
\end{abstract}




\section{Introduction}
\label{sec:intro}

Some giant planets on wide orbits around young stars can be directly detected because they are self-luminous and sufficiently bright relative to the host star at infrared wavelengths.  This method of detection can extend exoplanet surveys beyond the few au reached by the radial velocity (RV) technique \citep{Nielsen2019}.  Direct observations of giant planets can test formation models, e.g. core accretion vs. disk fragmentation, or thermodynamic ``hot start" vs. ``cold start" \citep{Spiegel2012,Marleau2014}, probe the composition of the disks from which the planets accreted gas \citep{Cridland2019,Zhang2020,Notsu2020}, constrain the dynamics of giant planet atmospheres, and reveal  circumplanetary disks \citep{Zhu2016,Wu2020}.   Understanding giant planet demographics is also important for studies of protoplanetary disk evolution, since these objects could be responsible for the formation of inner cavities \citep{vanderMarel2018} and warps in disks\citep{Nealon2018}.   Although the potential scientific gain is great, and instruments have steadily become more sensitive, giant planets on widely separated orbits appear rare \citep{Nielsen2019}; only about two dozen companions with masses less than the 13\mjup\ deuterium burning limit have been discovered in this manner \footnote{\url{https://exoplanetarchive.ipac.caltech.edu/}}.

We report a substellar companion to an M dwarf member of the 1-5 Myr-old Taurus star-forming region.  The host star, 2MASS~J04372171+2651014, hereafter referred to as 2M0437, was previously identified as a low-mass member of Taurus based on \spitzer/IRAC and 2MASS-\gaia\ proper motions and follow-up spectroscopy \citep{Esplin2017}.  The star was observed by the \kepler\ telescope during Campaign 13 of the \ktwo\ mission (8 March to 27 May 2017).  Like a significant fraction of Taurus stars \citep{Luhman2010}, 2M0437 lacks detectable excess infrared emission from a disk (Sec. \ref{sec:parameters}); nevertheless dust appears to orbit close to the star, partially occulting it, and manifesting itself as quasi-periodic dimming in the \ktwo\ light curve (Sec. \ref{sec:rotation}).  

Adaptive optics (AO) imaging in $H$-band ($\lambda = 1.6\mu$m) with the Subaru telescope in March 2018 revealed three significantly fainter sources in the neighborhood of 2M0437 (Fig. \ref{fig:field}).  Based on the statistics of such detections, the closet (0\arcsec.9 separation), which we designate 2M0437b, was considered as a candidate companion.  Here we report on our photometric, spectroscopic, and astrometric observations and analysis of the star and its faint companion, which demonstrate conclusively that 2M0437b is bound to the primary and is likely to have a super-Jupiter mass.  We also report a second object at a much wider (76\arcsec) which may be a bound stellar companion.      

\section{Observations and Data Reduction}
\label{sec:observations}

\subsection{AO Imaging with Subaru/IRCS}
\label{sec:ircs}

\begin{figure}
	\includegraphics[width=\columnwidth]{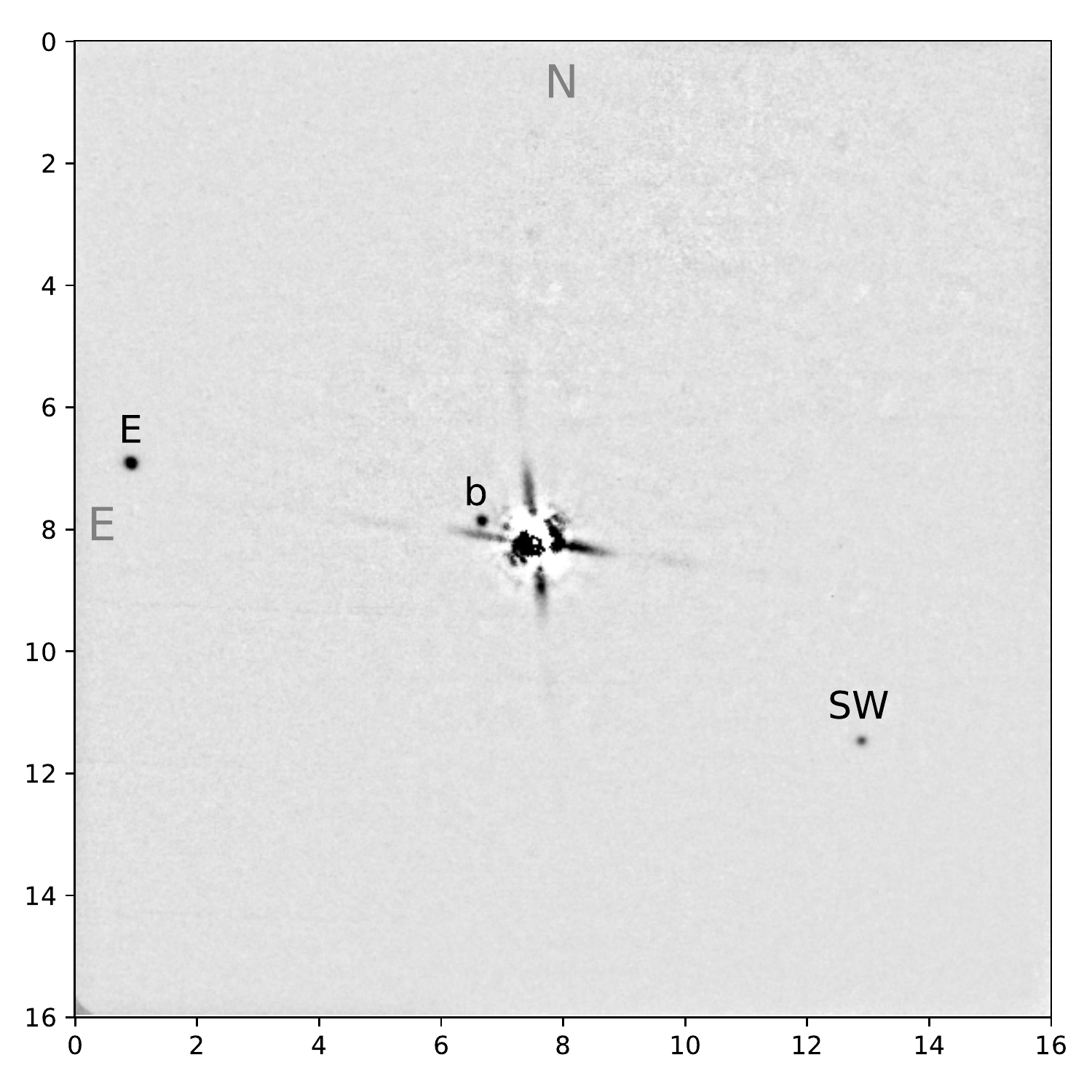}
    \caption{A $16 \times 16$ arc-sec portion of image of the 2M0437 field obtained through an $H$-band filter with the IRCS camera and AO 188 system on the Subaru telescope on UT 29 March 2018.  The image has had a radial median subtracted and subsequently a $65\times65$-pixel median filter applied.  "b" is the newly discovered companion, while "E" and "SW" are two background stars (see text).}  
    \label{fig:field}
\end{figure}

We performed AO imaging of 2M0437 with the InfraRed Camera and Spectrograph \citep[IRCS;][]{Kobayashi2000} and AO 188 system \citep{Hayano2010} on the Subaru 8.2-m telescope during the nights of UT 29 March 2018 and UT 2 April 2021. We used the fine-sampling mode of the IRCS camera ($1\,\mathrm{pix}\,\approx\,20$ mas) to achieve sufficient angular resolution, and took both unsaturated and mildly saturated frames to reach adequate dynamic range to measure the contrast between 2M0437 and neighboring sources.  Observations were carried out with five-point dithering using the $H-$band filter and $K^\prime-$band filter for the 2018 and 2021 runs, respectively. The integration times for each dithering position were set to 2 sec and $6-36$ sec for the unsaturated and saturated frames, respectively. We reduced the IRCS frames following \citet{Hirano2016}; dark-subtraction, bad-pixel masking, flat-fielding, and distortion-correction were applied to each frame, and we combined the aligned and co-registered frames for each of the unsaturated and saturated data sets.  The full width at half maximum (FWHM) of the combined unsaturated image was $\approx 0\farcs13$.  The combined saturated image taken in the 2018 run (in the $H-$band) exhibited three faint sources near 2M0437, separated by $\approx 0\farcs9$ (P.A. $\approx$ 70 deg.), $\approx6^{\prime\prime}$ (P.A. $\approx240$ deg.), and $\approx7^{\prime\prime}$ (P.A. $\approx80$ deg.) (``b", ``SW", and ``E", respectively in Fig. \ref{fig:field}).  Two of these sources (``b" and ``SW") were also captured in the combined image of the 2021 run. 

Astrometry was performed for the three faint objects in the combined IRCS images.  After shifting all saturated point spread functions (PSFs) of the primary star to a common center, we performed PSF subtractions. We calculated median values of the primary PSFs as a function of radius from the center and interpolated those to infer 1D radial profiles, which were converted to the 2D profiles and subtracted from the images. The positions of the faint source "b" in the radial-profile subtracted images were determined by fitting elliptical 2D-Gaussian functions to the PSFs of "b".  We also fit an elliptical 2D-Gaussian function to an unsaturated PSF of the primary star to estimate the position of the primary.   In order to determine the relative positional shifts between the saturated and the unsaturated PSFs of the primary, we compared them by performing least $\chi^2$ fitting, where the masks covering the saturated areas were applied to both the unsaturated and the saturated PSFs.   Accordingly, we estimated the $x$ and $y$ separations between the primary and the companion from all the saturated images, and computed the mean values and the standard deviations of those. For the other two objects (``E" and ``SW"), we estimated the relative positions in a similar fashion, but without subtracting the primary’s PSF.  Finally, the pixel $x$ and $y$ positions of the primary and each faint object were converted to the angular separation and position angle.   

\subsection{AO Imaging with Keck/NIRC2}
\label{sec:nirc2}

2M0437 was also observed with the NIRC2 infrared imager and Altair AO system on the Keck-2 telescope on the nights of UT 22 August 2018, 25 November 2018, 16 October 2019, 9 December 2019, and 4 January 2021. All observations used laser guide star adaptive optics, and the primary star served as its own reference star for tip-tilt correction. The images were obtained with the narrow camera, which gives a pixel scale of 9.971 mas pix$^{-1}$ \citep{Service2016} and a field of view of 10\farcs2.  All epochs used the $K'$ filter ($\lambda = 2.124$\micron) or $K$ filter ($\lambda = 2.196$\micron) since these offered the best AO correction and sensitivity to red companions. One epoch also used the $H$ ($\lambda = 1.633 \mu$m) and $J$ ($\lambda = 1.248 \mu$m) filters, though no sources besides the primary were detected in the observations at $J$-band due to poor wavefront correction and low Strehl ratio. A range of exposure times were used to obtain both unsaturated images of the primary star as well as deep imaging of the candidate companion; in most images, the companion was only detected if the primary star was saturated. The two sources at much wider separations were also observed in different subsets of the observations. We summarize the observations of 2M0437 that were taken in each epoch in Table \ref{tab:TabNIRC2}.

Our data analysis broadly follows the procedures outlined in \citet{Kraus2016}. To summarize, for each image we subtracted the nearest contemporaneous mode-matched dark frame with identical integration time, co-adds, and Fowler sampling, applied a non-linearity correction, and divided by the nearest contemporaneous flat field. We also performed ``de-striping'' in order to rectify spatially correlated read noise that is mirrored between the quadrants of the NIRC2 detector and that otherwise would have dominated the photometric noise budget for faint sources. Finally, we flagged all pixels that were saturated, impacted by cosmic rays, or known to be hot or dead. For pixels far from the location of the primary star, we used bi-linear interpolation to estimate and replace the missing pixel values. Bi-linear interpolation would not be appropriate in the saturated core of the primary star, so we instead replaced those pixel values with the scaled PSF of the best-fitting PSF reference template. However, we emphasize that these pixel values were only used for procedures requiring aperture photometry (such as detection limit determination) and for cosmetic display; our PSF-fitting results masked all bad pixels with NaNs and disregarded them in the fitting process.

To measure relative astrometry and photometry for all faint sources with respect to the primary star, we analyzed each science frame with our custom iterative PSF-fitting pipeline. The first stage of our pipeline adopts an initial estimate of the projected separation $\rho$, position angle $\theta$, and contrast $\Delta m$ for each faint source, and then tests the $\chi^2$ goodness of fit in matching the scene of the science frame with shifted/scaled copies of potential empirical PSF templates. We drew the template library from the 1000 unsaturated single-star observations in the public archive that are most contemporaneous with each science frame, windowing each potential template with a higher-order or ``super" Gaussian function just beyond its speckle halo, and we optimized the PSF match within a radius of 15 pixels around each source (or for the primary star, to a radius 10 pixels beyond the widest saturated pixel). The second stage of our pipeline then performs a $\chi^2$ minimization of $\rho$, $\theta$, and $\Delta m$ using that template, optimizing the relative astrometry and photometry for each source. The pipeline then iterates between these stages until the same best-fit template yields the lowest $\chi^2$ minimum in two consecutive iterations. Finally, we corrected the relative astrometry for the known optical distortion of NIRC2 using the distortion solution of \citet{Service2016}, which yields astrometry with a systematic noise floor of $\sim$1 mas for each source's position, or $\sim$1.4 mas in relative position between two sources. For each epoch, we adopt the average $\rho$, $\theta$, and $\Delta m$ for all observations taken in a given filter where the faint source was well detected, adopting the standard error as the statistical uncertainty in each measurement. We then add these standard errors in quadrature with the systematic uncertainties for relative astrometry ($\sim$1.4 mas) and photometry ($\sim$0.02 mag) to determine final uncertainties. 

To estimate the companion detection limits in each epoch, we also follow the procedure described in \citet{Kraus2016} using two different PSF-subtraction techniques. To probe wide separations that are limited by read noise and the smooth wings of the primary PSF, we produced a residuals map for each image by subtracting the azimuthally-averaged median profile of the primary star. To probe close separations that are limited by the speckles of the primary PSF, we also produced another residuals map for each image by subtracting the best-fitting empirical template that resulted from the astrometric and photometric analysis. We then produced significance maps for each case by measuring the RMS of the aperture photometry values in concentric annuli, normalized with respect to the aperture flux of the primary star, and stacked those maps (weighted by Strehl) to compute a final significance map centered on the primary star. Finally, we estimated the detection limit as a function of projected separation to be the contrast values corresponding to +6$\sigma$ outliers at a range of radii in that stacked significance map. We adopted these values from the azimuthal-median PSF subtraction for wide separations ($> 0\farcs.5$) and from the empirical PSF subtraction for close separations ($< 0\farcs.5$), setting the transition point according to the PSF method that yielded a deeper limit. Finally, to estimate the corresponding detection limits in terms of companion mass and projected physical separation, we adopted the Gaia EDR3 parallactic distance to 2M0437 and converted the limiting absolute magnitudes to masses using the 1 Myr and 5 Myr DUSTY models of \citet{Chabrier2000}.

\subsection{Moderate-low resolution spectroscopy}
\label{sec:modresspec}

A moderate resolution ($R\approx1000$) optical (3200-9700\AA) spectrum of the host star was obtained on UT 31 March 2018 with the SuperNova Integral Field Spectrograph \citep[SNIFS][]{Aldering2002,Lantz2004} on the UH~2.2m telescope on Maunakea.  The integration time was 920 sec and the observation airmass was 2.37.  Details of the spectrograph and data reduction are given in \citet{Gaidos2014a} and \citet{Mann2015}.  A 920 sec-spectrum of the DA white dwarf GD71 was also obtained at an airmass of 1.99 for telluric comparison.    

Near-infrared ($JHK$, 0.7-2.55\micron) spectra of 2M0437 were obtained on UT 19 and 25 April 2018 with the SpeX spectrograph on the NASA IRTF on Maunakea \citep{Rayner2004}.  The instrument was operated in short-wavelength, cross-dispersed (SXD) mode with a 0".3 slit and resolution of $\lambda/\Delta \lambda = 2000$.  Integration times were 120 sec, observation were obtained at an airmass of 1.84 and 2.37, respectively, and conditions were variable cloudiness with 1-2.5 mags extinction and photometric, respectively.  A low-resolution ($\lambda/\Delta \lambda =\approx 75$) $JHK$ (0.7-2.5 $\mu$m) spectrum of the second, wide-separation co-moving companion were obtained with SpeX in prism mode and a 0".8 slit  on UT 20 August 2020.    Conditions were photometric and an airmass varying between 1.44 and 1.1.  A spectrum of an A0-type star was obtained at an airmass of 1.08 for telluric correction.  Extraction and calibration and combination of all spectra were performed using the {\tt SpeXTool} package \citep{Cushing2004} and corrected for telluric absorption using the spectrum of an A0 star as described in \citet{Vacca2003}. 

\subsection{Echelle spectroscopy}
\label{sec:ird}

To measure the RV of 2M0437 to high precision, we obtained near infrared, high-resolution echelle spectra using the InfraRed Doppler (IRD) spectrograph \citep{Tamura2012,Kotani2018} on the Subaru 8.2-m telescope. From 2018 August through to 2020 February, a total of 32 IRD spectra were obtained with simultaneous reference spectra of the laser-frequency comb (LFC), with integration times of 600-1200 sec.  Each IRD spectrum covers the spectral region between 950\,nm and 1730\,nm, and the signal-to-noise (S/N) ratio was typically $\approx$35 per pixel at 1000\,nm. 

Raw IRD data were reduced in a standard manner, and we extracted one-dimensional stellar and LFC spectra individually. Based on those reduced spectra, precise RVs were computed using a dedicated analysis pipeline \citep{Hirano2020b}. In extracting the template spectrum of 2M0437 for the RV analysis, we combined multiple observed spectra with relatively high signal-to-noise ratios after correcting for the telluric features by fitting a model spectrum or using a telluric standard star observed on the same night.  Since the rotational velocity of the star is relatively large ($v\sin i>20$ km s$^{-1}$, see Sec. \ref{sec:rotation}), we optimized the length of each spectral segment for the RV analysis in order to cover multiple lines in each segment.  The resulting RVs have typical uncertainties of $30-60$ m s$^{-1}$ for each frame.  

\begin{figure}
	\includegraphics[width=\columnwidth]{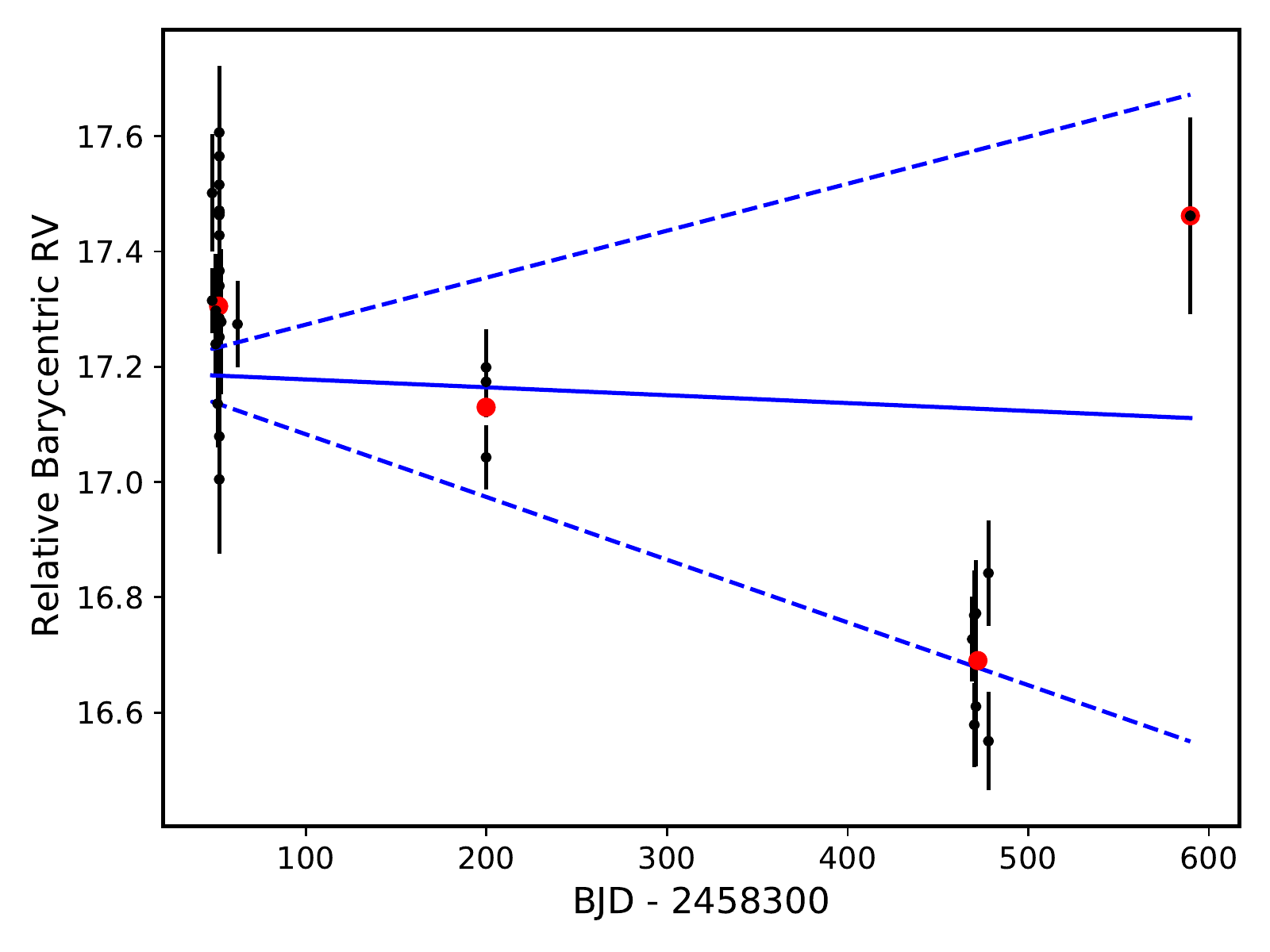}
    \caption{``Relative" RVs of 2M0437 obtained with the IRD spectrograph.  Black points are individual measurements while the red points are the means of each clump.  ``Relative" RVs are derived using more rigorous rejection of wavelength sections of spectra, and are more accurate in a relative (but not absolute) sense.  These RVs are based on the $YJ$-band portions of the spectra, which are more immune to the ``persistence" problems that affect Hawaii 2RG detectors.  The standard deviation of the means is 0.3 km sec$^{-1}$, and the dashed blue lines represent the 1$\sigma$ limits on the acceleration of the star.}  
    \label{fig:ird}
\end{figure}

We determined an "absolute" weighted mean barycentric RV of $17.1 \pm 0.3$ km sec$^{-1}$ using all segments of all epochs.  We then determined "relative" RVs which are more accurate in a relative sense, but not absolute sense, using only selected, well-behaved wavelength sections.  These are plotted in Fig. \ref{fig:ird}, after subtraction of the mean and addition of the "absolute" mean value.  Our IRD observations are highly clustered in time, and the scatter of the individual measurements around each mean is large compared to the formal errors, probably due to the effect of spots and stellar activity on this young star.  The individual means of each clump have a scatter of about 0.3 km sec$^{-1}$.  Assuming this is the true error, then a slope (acceleration) of up to $\pm$0.35 km sec$^{-1}$ yr$^{-1}$ is allowed by the observations and their scatter (dashed blue lines in Fig. \ref{fig:ird}).  

\subsection{\ktwo\ photometry}

\begin{figure*}
	\includegraphics[width=\textwidth]{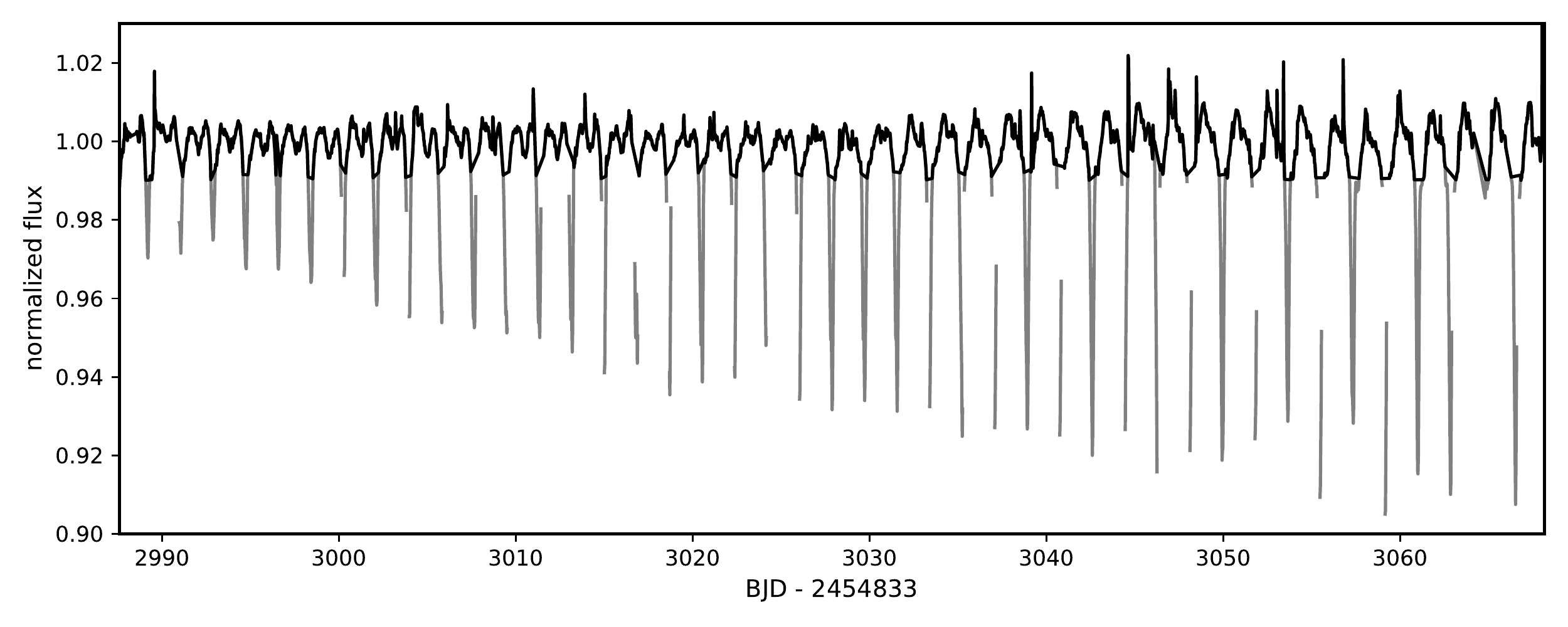}
    \caption{Normalized \ktwo\ PDCSAP-detrended light curve of 2M0437 obtained during Campaign 13.  Dimming events are interpreted to be from occultation by orbiting dust, and brightening events are flares.  Grey curves represent parts of the light curve that were excised when estimating a stellar rotation period.}  
    \label{fig:lightcurve}
\end{figure*}

Rotation periods of young stars can be routinely determined by precision photometry, especially from space.  2M0437, aka EPIC\,248131102 from the Ecliptic Plane Input Catalog of \ktwo\ \citep{Huber2016}, was observed by the \kepler\ telescope during Campaign 13 of the \ktwo\ mission (8 March to 27 May 2017).  The de-trended Pre-search Data Conditioning Simple Aperture Photometry (PDCSAP) \ktwo\ light curve contains a quasi-periodic signal with a period of 1.84 days (Fig. \ref{fig:lightcurve}).  This signal is also seen in light curves constructed using Simple Aperture Photometry (SAP), and the {\tt k2sff} and {\tt everest} pipelines \citep{Vanderburg2014,Luger2016} and so is not an artifact of the reduction.  Instead, this dimming is characteristic of some ``dipper" stars with close-in orbiting dust, although these stars usually have primordial disks \citep[e.g.,][]{Ansdell2016b,Stauffer2017,Cody2018,Bredall2020}.

\section{Results and Analysis}
\label{sec:analysis}

\subsection{Primary Star Properties}

\subsubsection{Taurus membership}

\begin{figure}
	\includegraphics[width=\columnwidth]{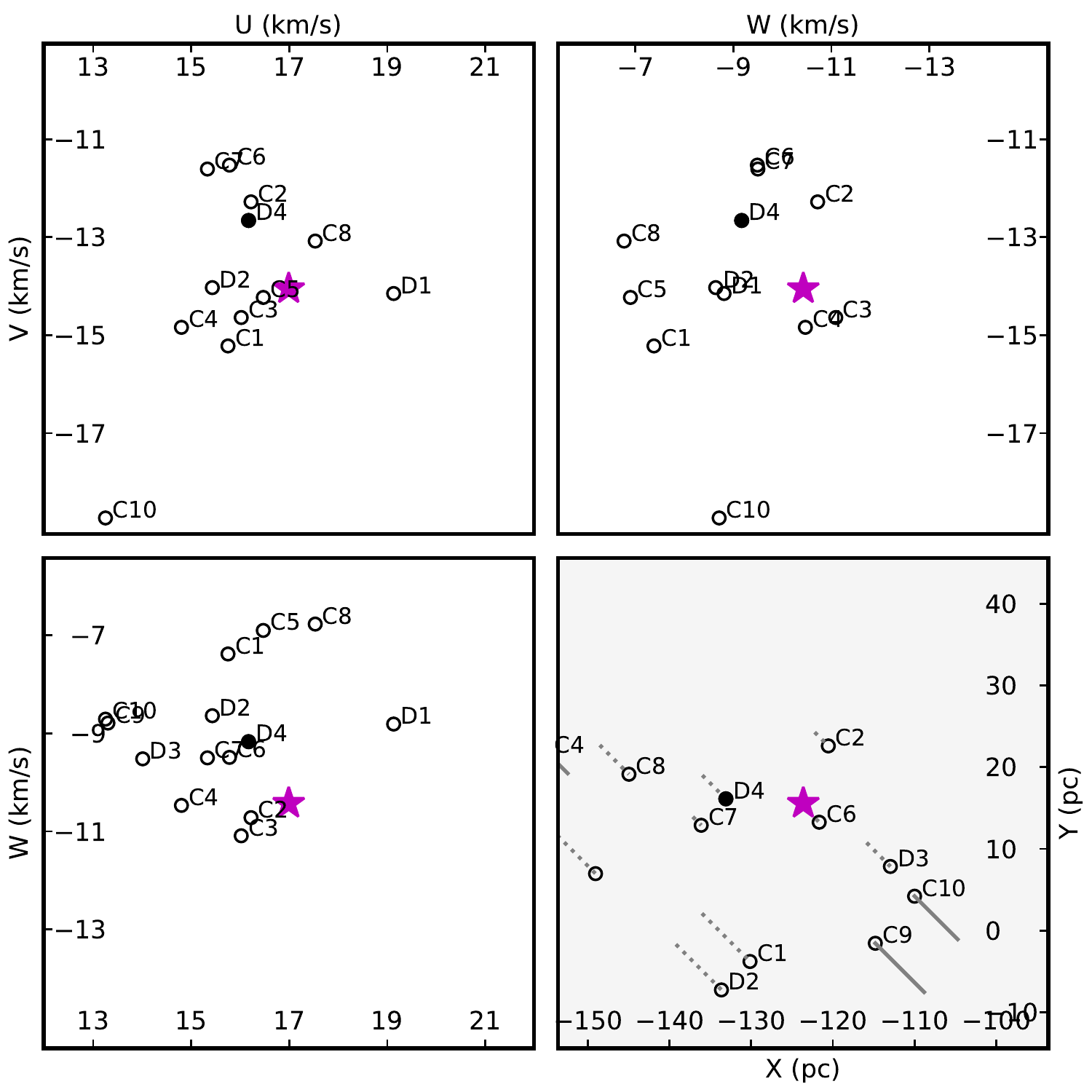}
    \caption{Upper left and right and lower left panels: $UVW$ space motion of 2M0437 (magenta star) relative to some of the Taurus groups cataloged by \citet{Krolikowski2021} (open circles).  We identify the dispersed D4 group (filled circle) as the most likely host group.  Lower right: Position of 2M0437 relative to those groups in Galactic Cartesian coordinates.  $Z$ values (relative to 2M0437) are represented as the length of "pins".  Coordinate systems are left-handed, i.e. positive $U$ and $X$ towards the Galactic center.}  
    \label{fig:uvwxyz}
\end{figure}

The Galactic position and space motion of 2M0437 was calculated using \gaia\ astrometry and the IRD-based RV (Table \ref{tab:primary}).  The $UVW$ space motion (left-handed coordinate system) of 2M0437 is ($+17.0 \pm 0.3$, $-14.05 \pm 0.07$, $-10.43 \pm 0.08$)\,\ks\ and unambiguously associates it with the Taurus cloud (Fig. \ref{fig:uvwxyz}).  However, Taurus has long been known to be structured and heterogeneous \citep[e.g.,][]{Lynds1962,Gomez1993}, and that structure has been clarified with expanded membership catalogs and precise astrometry from \gaia. \citet{Luhman2018} used \gaia\ DR2 data to describe multiple spatial and kinematic clusters associated with individual clouds. Among these, the $UVW$ of 2M0437 is closest to that of the L1517 group and to a lesser extent the adjacent L1527 group, however the position on the sky favors L1527.   \citet{LiuJ2021} performed density-based clustering with the {\tt DBSCAN} algorithm applied to Taurus candidates in a 5-dimension position and proper motion space.  They identified 8 ``young" groups with ages of 2-4 Myr based on a comparison to {\tt PARSEC} isochrones \citep{Bressan2012}, plus 14 older (8-11 Myr) groups.  2M0437 was assigned to the first of the 8 young groups, which also corresponds to Group A of \citet{Roccatagliata2020}.  Finally, \citet{Krolikowski2021} performed a clustering using the latest (EDR3) \gaia\ astrometry and identified 17 groups; ranked by kinematic similarity, 2M0437 is most closely affiliated with the C3 and C4, the "halo" and "core" of L1517 group of \citet{Luhman2018}.  However, those groups are 30 pc away and the closest group in space is C6 (L1524), which itself is a poor kinematic match.  The actual best match is not clear, but the group which is highest in both kinematic and spatial ranking is C2 (L1495) followed by D4 North.  The separation of 2M0437 from the centers of C2 and D4 (10.4 pc and 16.0 pc) can be better explained by the distributed spatial scattering of D4 than that of the concentrated C2 group.  Thus we tentatively assign 2M0437 to D4, for which \citet{Krolikowski2021} estimate an age of $2.5 \pm 0.35$ Myr.

\subsubsection{Stellar Parameters}
\label{sec:parameters}

We consider 2M0437 to be a single star.  As discussed in Section~\ref{sec:limits}, our AO observations show no stellar companions to within $\approx$0\farcs1 ($\sim$13 AU) and no companion above the deuterium-burning limit to within $\approx$0\farcs15 ($\sim$20 AU).  The goodness of fit of \gaia\ astrometry to a single-star solution, as quantified by a Renormalized Unit Weight Error (RUWE) value is 1.214, falling within the range of single stars \citep{Kervella2019,Belokurov2020}, (but see \citealt{Stassun2021}).  The time-averaged absolute acceleration induced by a minimum hydrogen-burning mass (0.08\msun) companion on a highly-inclined circular orbit of semi-major axis $a$ around this $\approx$0.18\msun\ star  (see Sec. \ref{sec:parameters}) is 9.5 (a/1 AU)$^{-2}$ km sec$^{-1}$~yr$^{-1}$.  Thus at a two standard-deviation we used the RV data to reject potential stellar companions at $a<40$ AU, or a projected separation of 0\arcsec.3.  Thus, barring some exceptionally unfortunate orbital phase, we rule out any stellar-mass companions to 2M0437.  

The SNIFS and SpeX spectra were combined to form a single, discontinuous spectrum covering 0.32-2.4 \micron, using the overlap regions between the SNIFS blue and red channels and SpeX $YJ$-band to compute relative normalizations.  Synthetic magnitudes are calculated by integrating over pass-band response functions and comparing to published photometry:  $BVgri$ photometry was obtained from the APASS DR10 \citep{Henden2019}, \gaia\ $GB_pR_p$ magnitudes from \gaia\ EDR3 \citep{Gaia2020}, $JHK_s$ from 2MASS \citep{Skrutskie2006}, IRAC 4-band photometry from \spitzer\ \citep{Rebull2010}, and W1-W4 (2.4-25 $\mu$m) photometry from the AllWISE survey \citep{Cutri2013}.  The observed spectrum is flux-calibrated by matching the synthetic photometry to the observations, and interpolated models of PHOENIX stellar spectra \citep{Husser2013} are fit to the distorted spectrum.  Stellar atmosphere parameters (\teff, $\log g$, [Fe/H]) are then inferred from the best-fit model spectrum, and the bolometric flux is determined by integrating over the flux-calibrated spectrum, using the best-fit model to replace missing sections, and Rayleigh-Jeans extrapolation at wavelengths beyond $K$-band.  The spectral corrections are (1) an overall multiplicative factor; and (2) a wavelength- and air-mass dependent correction derived from a standard star observation.  

The best-fit interpolated model spectrum has a $\teff=3104K$, \logg\ = 3.93, and [Fe/H] = +0.01.  The derived metallicity agrees with that found overall for the Taurus-Aurigae star-forming region \citep[$0.01 \pm 0.05$;][]{DOrazi2011}.   We also performed a fit that includes wavelength-independent veiling \citep{Herczeg2014} term that is set to the mean of the difference between the model- and the flux-calibrated spectra, but this did not significantly affect our results.   

\begin{figure}
	\includegraphics[width=\columnwidth]{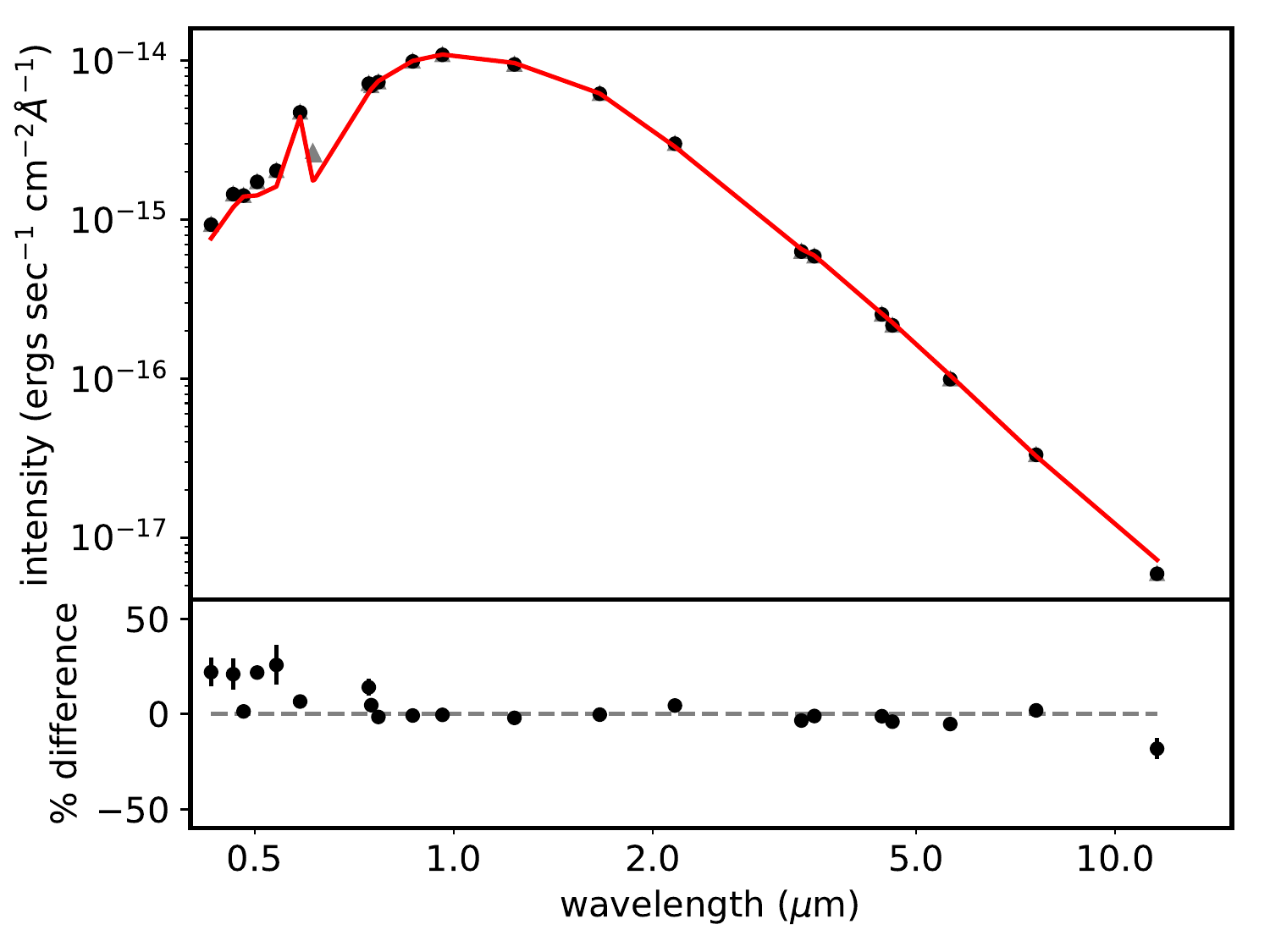}
    \caption{Spectral energy distribution of 2M0437 found by fitting a BTSETTL solar metallicity models with CIFIST opacities \citep{Baraffe2015} to available photometry (see text) within the Virtual Observatory's SED analyzer.  The discrepant $r$-band data (grey triangles) are excluded from the fit.  The best-fit model (red curve) has \teff=3100\,K and \logg=5.  The difference between the best-fit model and observations (lower panel) does not suggest any excess infrared emission related to a circumstellar disk.}  
    \label{fig:sed}
\end{figure}

We also fit the available photometry (Fig. \ref{fig:sed}) directly using the Virtual Observatory Spectral Energy Distribution analyzer \citep{Bayo2008} and the BT-Settl-CIFIST series of solar-metallicity stellar atmosphere models \citep{Baraffe2015}. Both $r$-band measurements (APASS and PanSTARRS) are significantly above the best-fit model predictions, possibly due to inaccuracies in modeling of deep molecular bands or \halpha\ emission from intermittent accretion (see Sec. \ref{sec:disk}), and were excluded from the fit.   The best fit ($\chi^2=128$, $\nu=14$) with zero reddening is for a 3100\,K and \logg = 5 model; a Bayesian analysis gives 96\% confidence that \logg\ is between 4.51 and 5.50.  The bolometric flux is $1.337 \pm 0.005 \times 10^{-10}$ ergs sec$^{-1}$ cm$^{-2}$.  Combined with a \gaia-based distance, the bolometric luminosity is then $0.0689 \pm 00005$ \lsun.  Adopting \teff=$3100\pm100$\,K, the radius is $0.84 \pm 0.11$\rsun, consistent with that of a greatly inflated pre-main sequence M dwarf.  We constrain the amount of reddening/extinction by varying \ebv\ and re-fitting; we find that for \ebv $\lesssim 0.1$, the best-fit \teff\ to vary from the spectroscopic value by more than 100\,K, nor $\chi^2$ to increase significantly.   
\begin{figure}
	\includegraphics[width=\columnwidth]{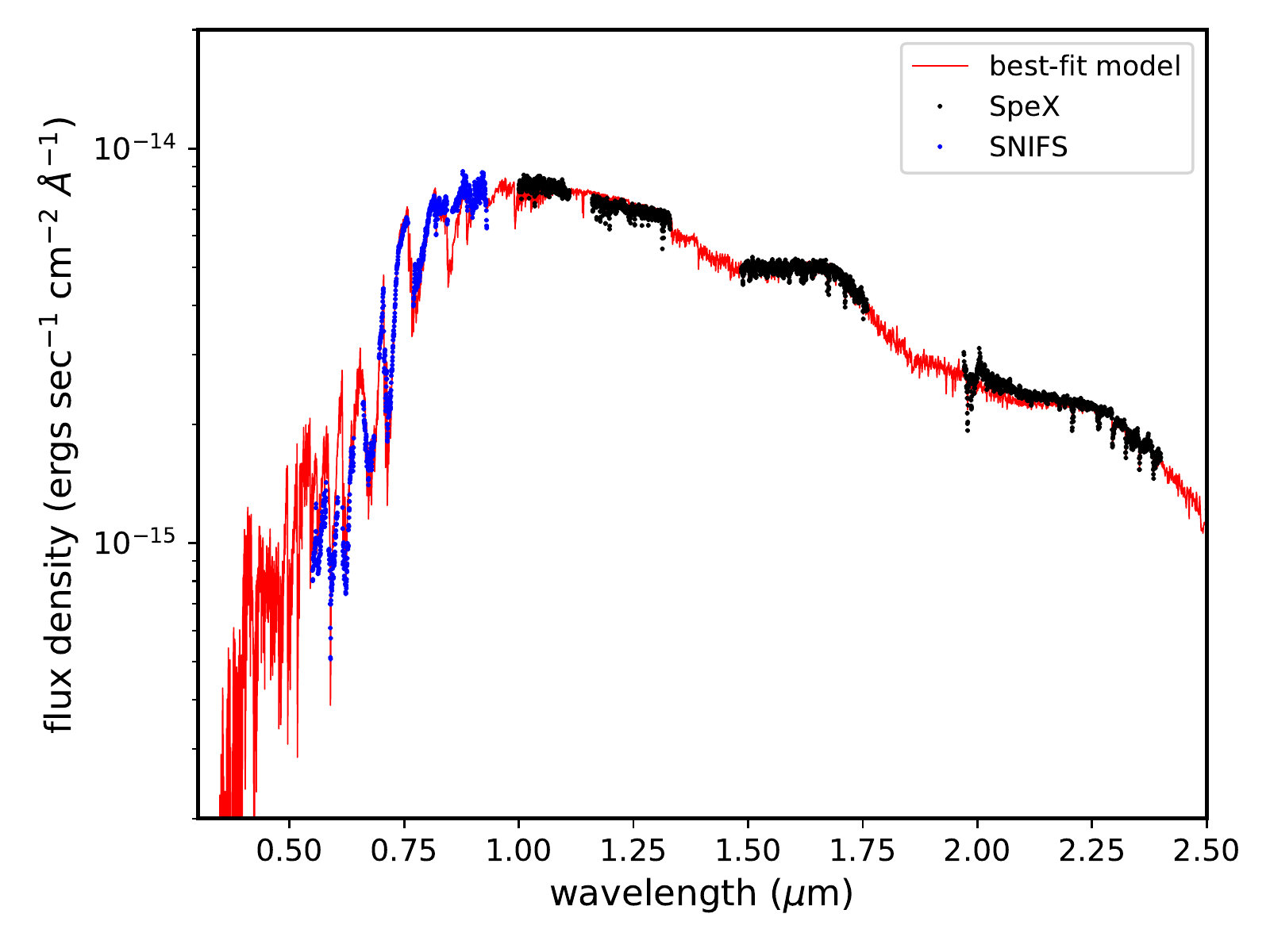}
    \caption{Combined SNIFS (visible, blue points) and SpeX (near-infrared, black points) spectra compared with a best-fit interpolated PHOENIX ACES model \citep{Husser2013}. The best fit model (red line) has \teff = 3104K, \logg = 3.93, and [Fe/H]=+0.01.}  
    \label{fig:spectrum}
\end{figure}

\subsubsection{Disk and Accretion}
\label{sec:disk}

\begin{figure}
	\includegraphics[width=\columnwidth]{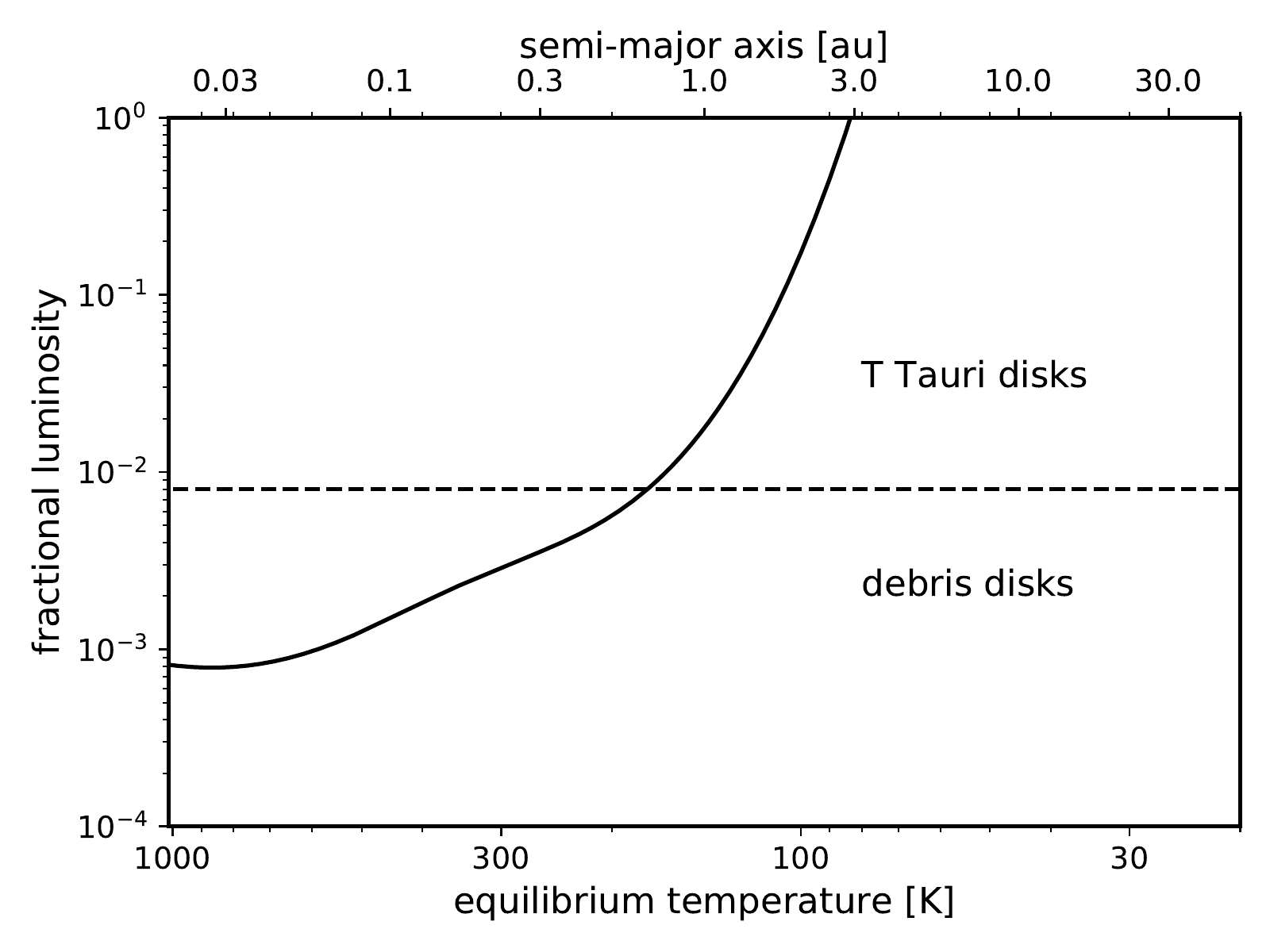}
    \caption{95\% confidence limits ($\Delta \chi^2 < 19.7$, 11 degrees of freedom) on the fractional luminosity of isothermal, black-body-like dust around 2M0437 based on the available photometry at $\lambda > 0.8$\micron.  The red dashed line is the debris-disk/accretion disk boundary suggested by \citet{Hughes2018}.}  
    \label{fig:disk}
\end{figure}

There is no detectable emission, i.e., due to a circumstellar disk, above the model through the \wise\ W3 (12 \micron) bandpass (Fig. \ref{fig:sed}).  Based on a differential $\chi^2$ analysis, the SED rules out classical T Tauri disks, but not "transition" disks, i.e. those with a cavity out to $\sim$3 au, and only excludes debris disks that are warm ($>200$K) and substantial (Fig. \ref{fig:disk}).   These weak limits are characteristic of observations of low-luminosity M dwarfs, which suffer from flux bias in surveys \citep{Avenhaus2012}.  The \halpha\ line was in emission with an equivalent width of 0.54\AA\ at the epoch of the SNIFS observation, consistent with moderately active M dwarfs \citep{Ansdell2015}, and well below the few \AA\ that is typical of weak-lined T Tauri stars, let alone classical T Tauri stars ($>20$\AA; \citealt{White2003}).

\subsubsection{Mass and Age}

\begin{figure}
	\includegraphics[width=\columnwidth]{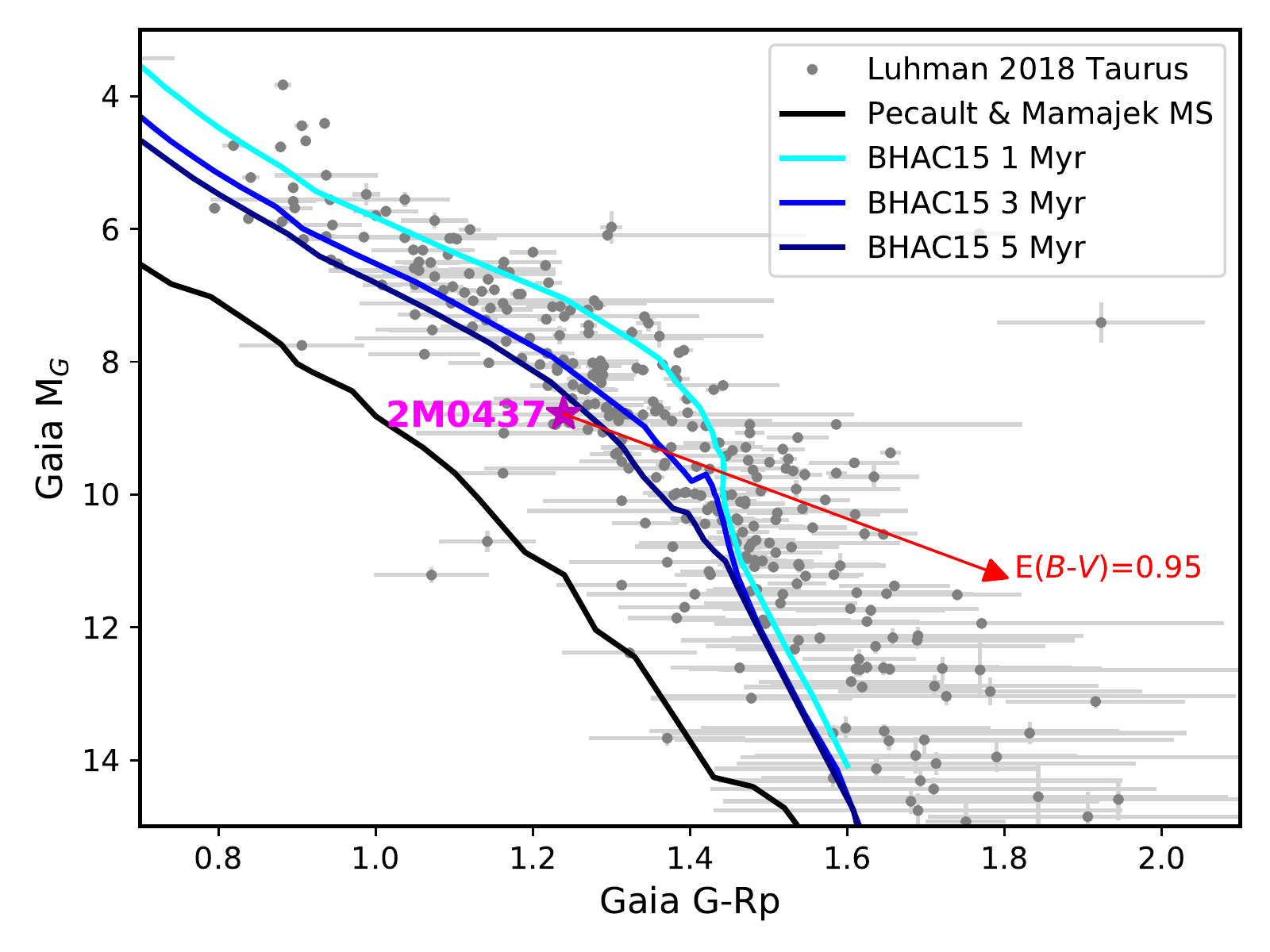}
    \caption{\gaia-based color-magnitude diagram of Taurus members cataloged by \citet{Luhman2018}, including 2M0437.  The empirical main sequence of \citet{Pecaut2013} and some pre-main sequence isochrones from the BHAC15 models of \citet{Baraffe2015} are plotted.  The reddening vector for \ebv=0.95 (the integrated reddening through Taurus along this line of sight) is based on the reddening coefficients for a \teff=5240\,K star from \citet{Casagrande2018}.  The location of 2M0437 on the blueward side of the locus may be a result of the lack of reddening along the line of sight to the star.}  
    \label{fig:cmd}
\end{figure}

The location of 2M0437 in a \gaia-based color-magnitude diagram (CMD)  is close to  the locus of Taurus stars (Fig. \ref{fig:cmd}); its location blueward of the observed locus of members is probably due in part to the negligible reddening along the line of sight compared to most members (Sec. \ref{sec:parameters}), perhaps because it is in front of the molecular cloud.  The star is also bluer than 1-5 Myr pre-main sequence isochrones \citep{Baraffe2015}, but predicted absolute \gaia\  magnitudes are derived from more fundamental model parameters via stellar atmosphere models and are probably not accurate to $<0.1$ mags.  Instead, we inferred the mass and age of the star, by comparing its derived \teff\ and bolometric luminosity to these sets of stellar evolution models, the \citet{Baraffe2015} models of pre-main sequence stars, the Dartmouth models with and without the expected effects of magnetic fields \citep{Dotter2008,Feiden2016}, and the SPOTS models of \citet{Somers2020}, which includes some of the expected effects of star spots.  These models are not entirely independent, since the SPOTS models use the \citet{Allard1997} model stellar atmospheres for M dwarfs and the \citet{Baraffe2015} models use a derivative of these with different line lists \citep{Rajpurohit2013}.  Magnetic activity and star spots are thought to significantly affect the properties and evolution of low-mass stars, particularly the luminosity and thus the inferred mass and age \citep[e.g.,][]{Feiden2016}.  2M0437 is rotationally variable with an amplitude of 1\%, at the lower end of the variability range of Taurus members \citep{Rebull2020}.  The expected spot fraction of $<10$~Myr-old stars is $\lesssim10$\% \citep{Morris2020}.   

\begin{figure}
	\includegraphics[width=\columnwidth]{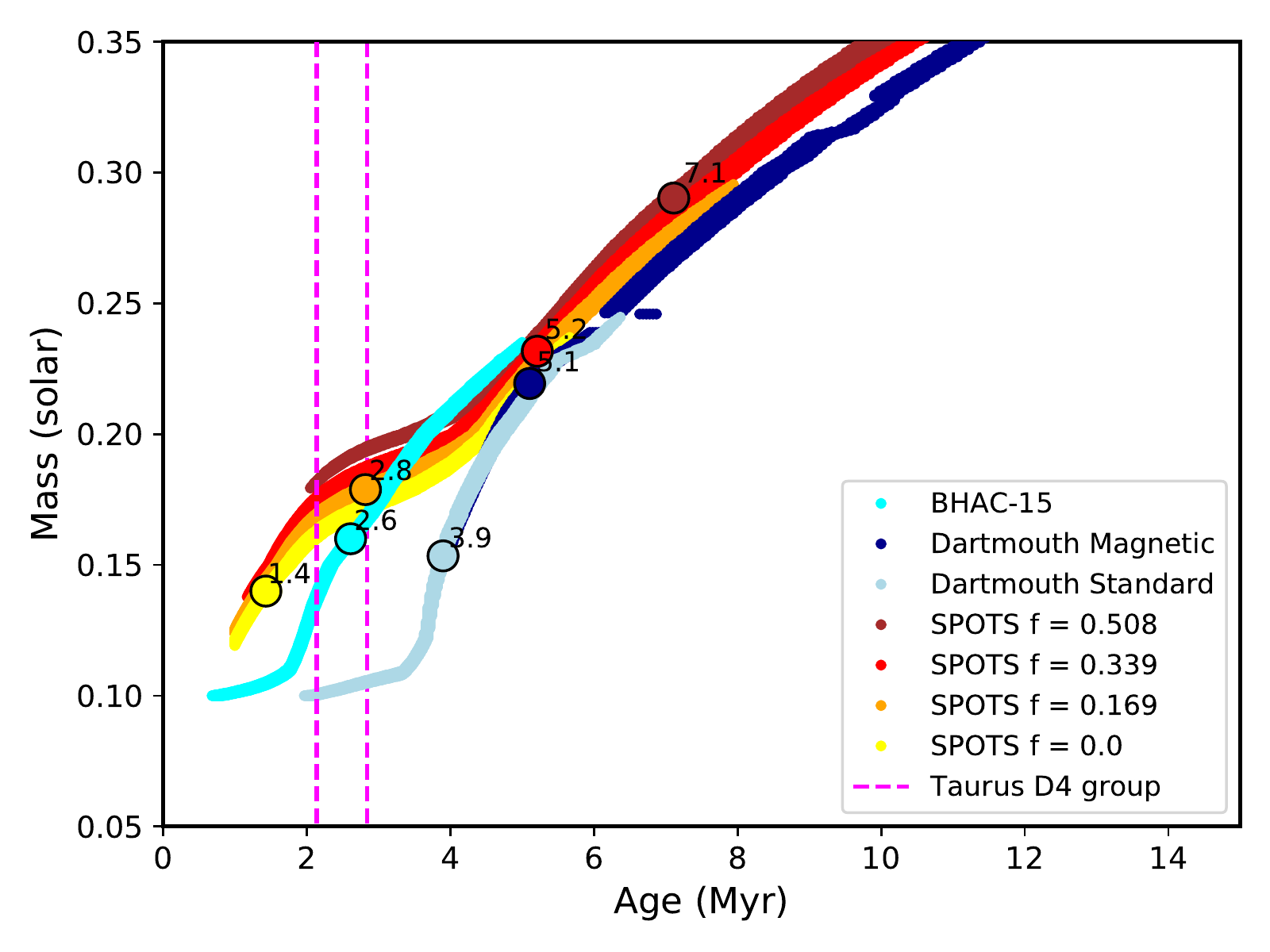}
    \caption{Masses and ages of BHAC-15 \citep{Baraffe2015}, Dartmouth standard and magnetic models \citep{Dotter2008,Feiden2016}, and SPOTS models \citep[][, with different spot fractions $f$]{Somers2020} that are consistent within a 95\% confidence level ($\chi^2 <6$ for 2 degrees of freedom) with the luminosity and \teff\ of 2M0437.  Each color represents those interpolated models with \teff\ and \lbol\ consistent with observations and with a specified spot coverage fraction.  The colored circles are the best-fit (minimum $\chi^2$) values, labeled with the age.}  
    \label{fig:mass-age}
\end{figure}

We linearly interpolated the published models onto a finer grid in $M_*$ and $\log$ age and computed the $\chi^2$ between the grid and the inferred values of \teff\ and $L_*$.  The uncertainties were calculated based on the distribution of $\chi^2$ values.  Figure \ref{fig:mass-age} shows the distribution of consistent models with stellar mass and age for BHAC15 and for SPOTS with spot coverage fractions up to 50\%.  Although best-fit BHAC15 models and low spot fraction SPOTS models have predicted ages $<$3 Myr that are roughly consistent with that of the D4 Taurus group of \citet{Krolikowski2021} (unsurprising since that is also based on similar models), mass and age for an individual star are highly covariant and thus poorly constrained.  For nominal age of $\approx 2.5$~Myr, based on the affiliation with D4 subgroup of Taurus \citep{Krolikowski2021}, the BHAC15 and SPOTS model-based mass ranges between 0.15 and 0.18\msun (Fig. \ref{fig:mass-age}), but we cannot rule out the possibility of an older age and higher mass if the star is much more magnetized/spotted than assumed. 

\subsubsection{Rotation}
\label{sec:rotation}

\begin{figure}
	\includegraphics[width=\columnwidth]{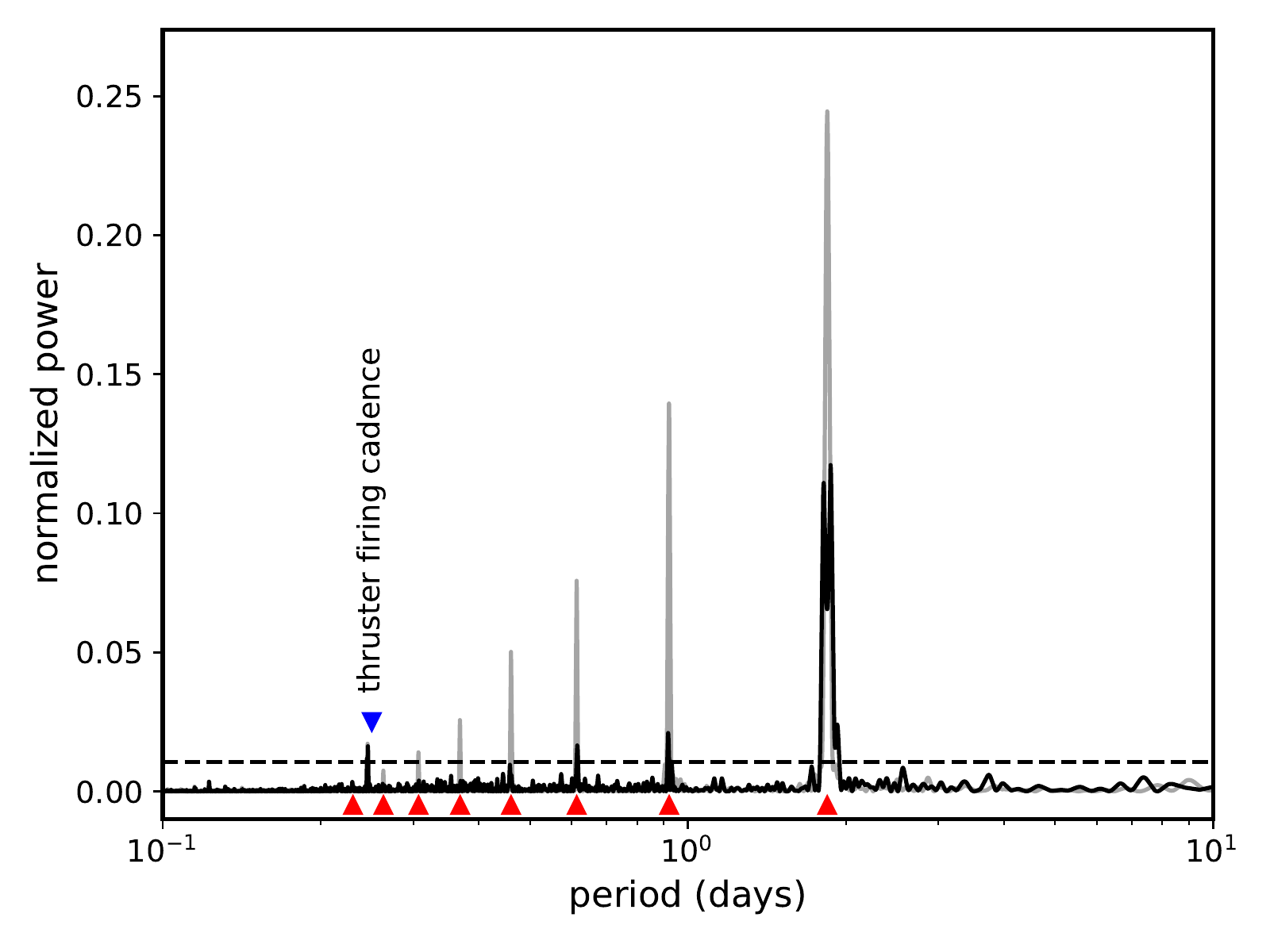}
    \caption{Lomb-Scargle periodogram of the \ktwo\ light curve of 2M0437.  The dashed line is the predicted power level for a false alarm probability of 1\%.  The 1.84 day period and its harmonics are marked by the red  triangles and the 6 hr typical cadence of reaction thruster firings is marked by the blue inverted triangles.  The grey curve is the power spectrum when the pronounced dimming events are excised from the light curve (grey portions in Fig. \ref{fig:lightcurve}.}
    \label{fig:periodogram}
\end{figure}

\begin{figure}
	\includegraphics[width=\columnwidth]{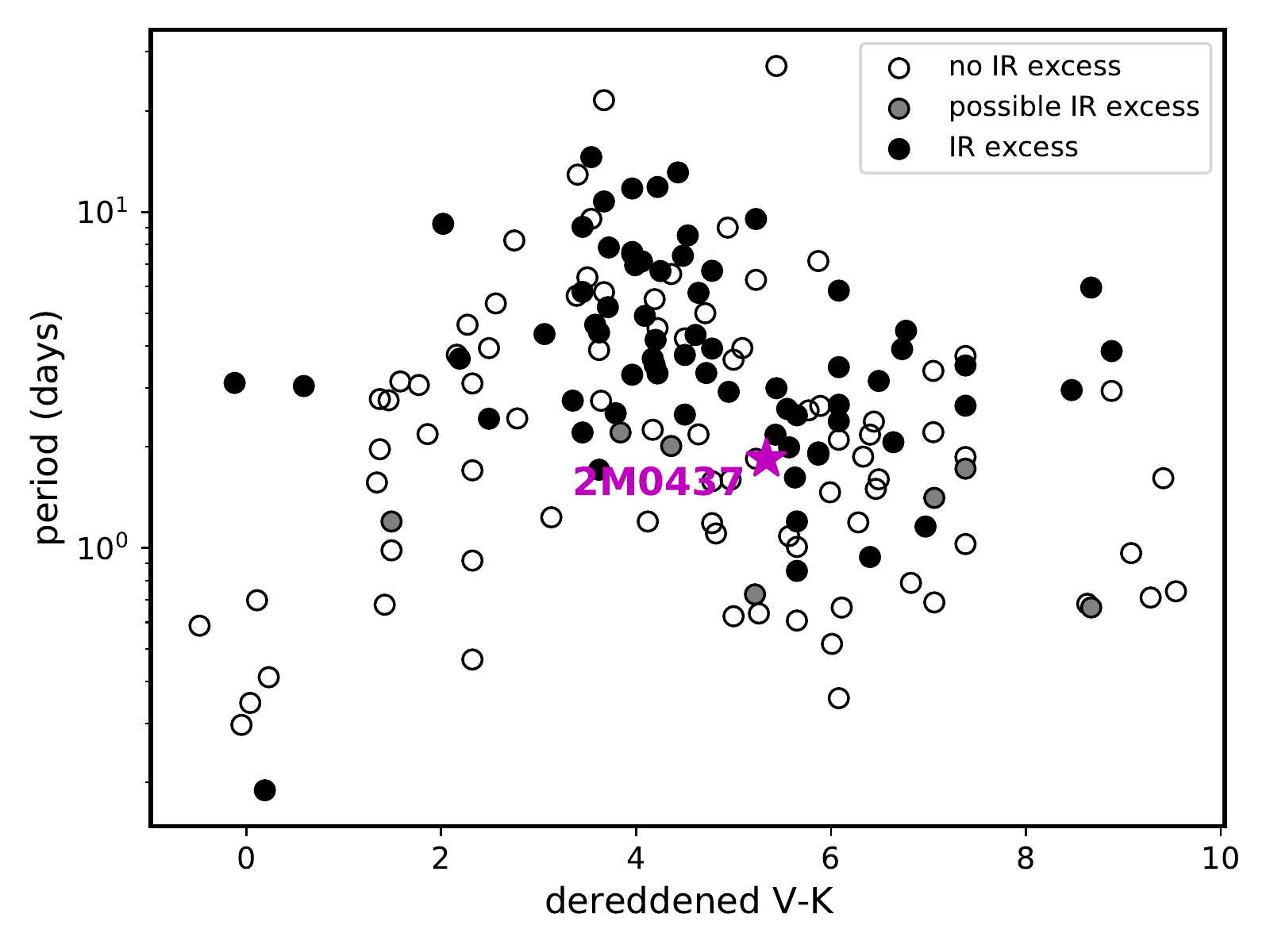}
    \caption{The rotation period of 2M0437 compared to those of Taurus members or candidate members from \citet{Rebull2020}.  Symbols are keyed according to the presence or absence of an infrared excess due to a circumstellar disk.}  
    \label{fig:periods}
\end{figure}

A Lomb-Scargle periodogram of the light curve from 0.1-10 days contains a single peak at 1.84 days plus upper harmonics produced by the quasi-periodic nature of the signal (Fig. \ref{fig:periodogram}). The only other significant peak is near 6 hours, the typical cadence of reaction thruster firing during the \ktwo\ mission.  To enhance the search for other signals, a periodogram was constructed from a clipped light curve, i.e. points below a threshold of 0.99 were excluded (grey domains in Fig. \ref{fig:lightcurve}).  This partly suppressed the 1.84-day peak (grey curve in Fig. \ref{fig:periodogram}) but did not reveal any additional peaks.  We surmise that the rotation period of the star is also near 1.84 days and that the dust is in a co-rotation orbit, as has been reported for some other "dipper" stars \citep{Stauffer2017}.  This rotation period is not uncharacteristic for cool Taurus stars (Fig. \ref{fig:periods}).

\begin{figure}
	\includegraphics[width=\columnwidth]{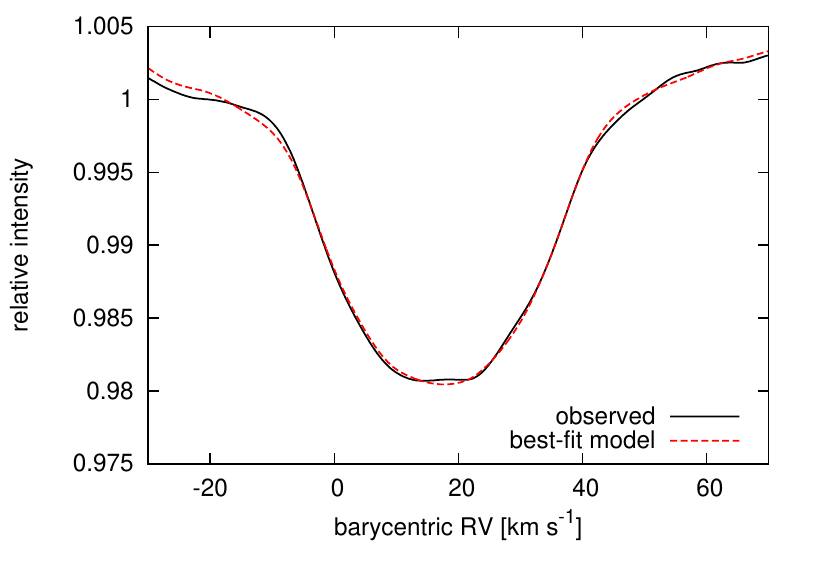}
    \caption{Normalized cross-correlation function (CCF) for 2M0437 calculated from the IRD spectra (black solid line). The best-fit theoretical CCF to the observed one is plotted as the red dashed line. }  
    \label{fig:ccf}
\end{figure}

By combining the rotation period with our estimate of the stellar radius (Sec. \ref{sec:parameters}) we derived an the equatorial rotation speed of $23 \pm 3$ km\,sec$^{-1}$. We compared this with the projected rotation velocity $v\sin i$ of the star derived from the IRD spectra. Following \citet{Hirano2020}, we cross-correlated 2M0437's IRD spectrum against the template IRD spectrum of GJ 699 (Barnard's star), having a similar surface temperature to 2M0437, and compared the cross-correlation function (CCF) with theoretical CCFs generated for various values of $v\sin i$; Convolving GJ 699's spectrum with the rotation plus macroturbulence broadening kernel \citep{Hirano2011} with $18.0\,\mathrm{km~sec^{-1}} \leq v\sin i\leq 27.0\,\mathrm{km~sec^{-1}}$ in steps of $0.5\,\mathrm{km~sec^{-1}}$, we created model IRD spectra, each of which was cross-correlated against the template as for 2M0437's observed spectra. The observed CCF of 2M0437 was fitted by MCMC by interpolating those CCF models for any given value of $v\sin i$ between $18.0\,\mathrm{km~sec^{-1}}$ and $27.0\,\mathrm{km~sec^{-1}}$. This analysis yielded $v\sin i=22.54 \pm 0.14$ km sec$^{-1}$; the observed and best-fit model CCFs are shown in Fig. \ref{fig:ccf}.  2M0437's $v\sin i$ is in remarkable agreement with the equatorial rotation velocity of the star, supporting a rotation period of 1.84 days and indicating  that the stellar inclination $i\approx 90^\circ$ (i.e., the stellar equator is nearly edge-on). 

\subsection{Background Stars}

Two fainter ($K \sim 17$) sources appear in our infrared imaging at $\rho = 6\arcsec.7, \theta = 78$\,deg (labeled "E" in Fig. \ref{fig:field}) and $\rho = 6\arcsec.2, \theta = 239$\,deg (labeled "SW").  These sources also appear in the UKIDSS Galactic Plane and Clusters surveys \citep{Lucas2013}, but, due to their faintness at visible wavelengths, not in \gaia\ EDR3, and thus absolute astrometry is not available to assess whether they are associated with 2M0437 or the Taurus cluster itself.  These objects are too faint to be cool giant stars for any reasonable level of extinction and are probably cool dwarfs that are either associated with Taurus or are in the background.  \footnote{Having two unaffiliated foreground objects within a few arcsec of any line of sight is statistically unlikely and the objects would have to be rare ultra-cool dwarfs.}  

\subsubsection{Astrometry}
\label{sec:back_astrometry}

\begin{figure*}
	\includegraphics[width=\textwidth]{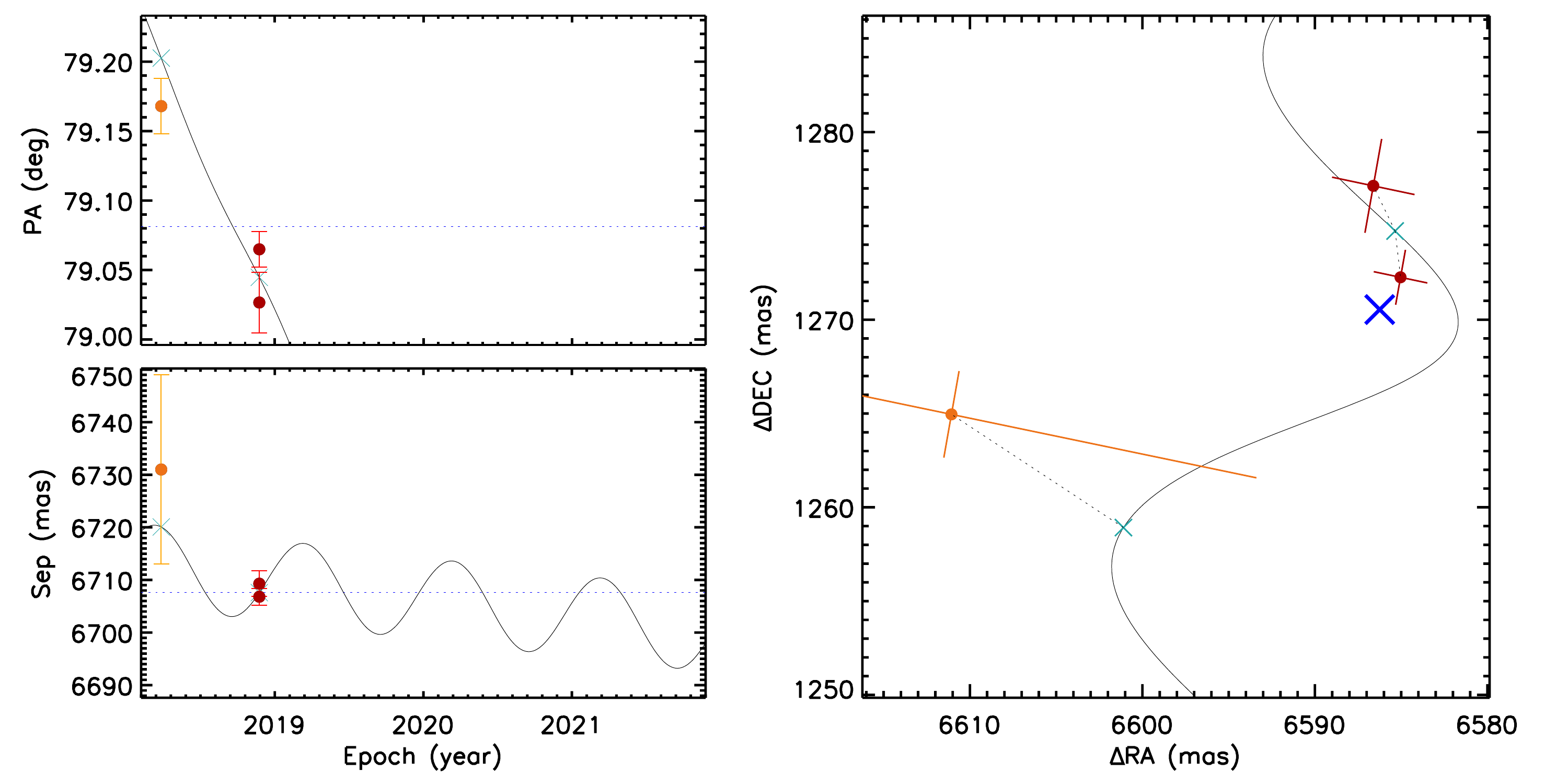}
	\includegraphics[width=\textwidth]{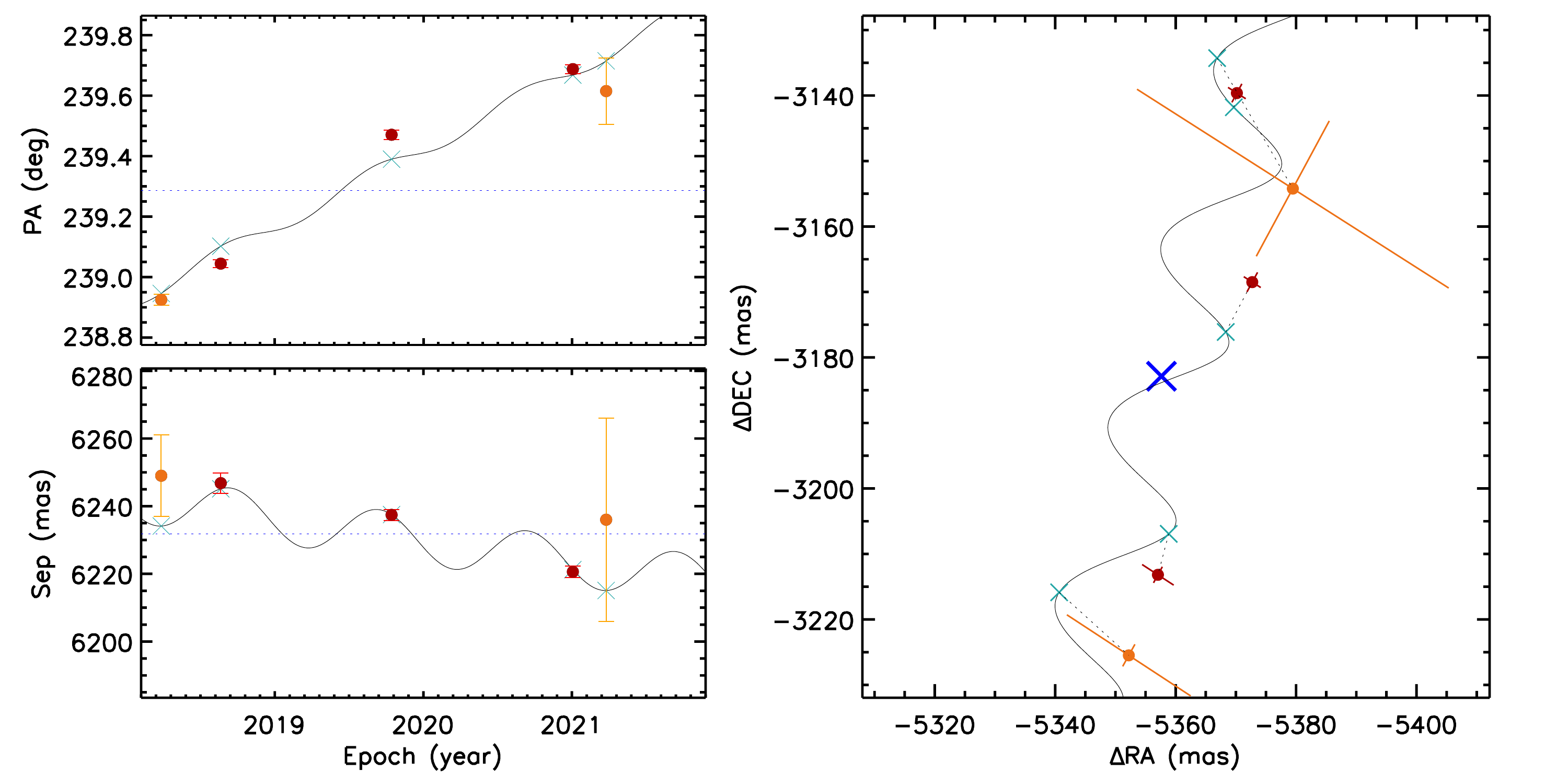}
    \caption{Astrometry of the E (top panels) and SW (bottom panels) background stars.  Position angle and separation are plotted on the left side, while R.A. and declination relative to 2M0437 in mas are plotted on the right.  The solid curves are the predictions for a star at the distance of Taurus that is motionless in the \gaia\ reference frame.  The dashed blue are the predictions for a hypothetical co-moving object.  Red points are NIRC2 and orange points are IRCS data.  A 0.25 deg offset has been applied to combine these datasets (see text).}  
    \label{fig:background_astrometry}
\end{figure*}

We obtained astrometry of the two wide-separation (E and SW) sources from our multi-epoch, combined NIRC2 and IRCS AO imaging.  To reconcile the two datasets, we introduce a rotational offset of 0.25 deg.  Our data rule out or disfavor common-proper motion of either source with 2M0437 (Fig.~\ref{fig:background_astrometry}).  The astrometry for the SW source is definitively inconsistent with a hypothesis of co-movement ($\chi_{\nu}^2 = 210$ with 8 degrees of freedom), but much more consistent with a stationary distant background star ($\chi_{\nu}^2 = 6.5$ with 8 degrees of freedom). The astrometry for the E source is broadly inconsistent with co-movement ($\chi_{\nu}^2 = 7.3$ with 4 degrees of freedom) and consistent with a stationary star ($\chi_{\nu}^2 = 1.8$ with 4 degrees of freedom). The $p$-value for the hypothesis of co-movement is $P < 0.001$.

\subsubsection{Photometry}
\label{sec:back_photometry}

\begin{figure}
	\includegraphics[width=\columnwidth]{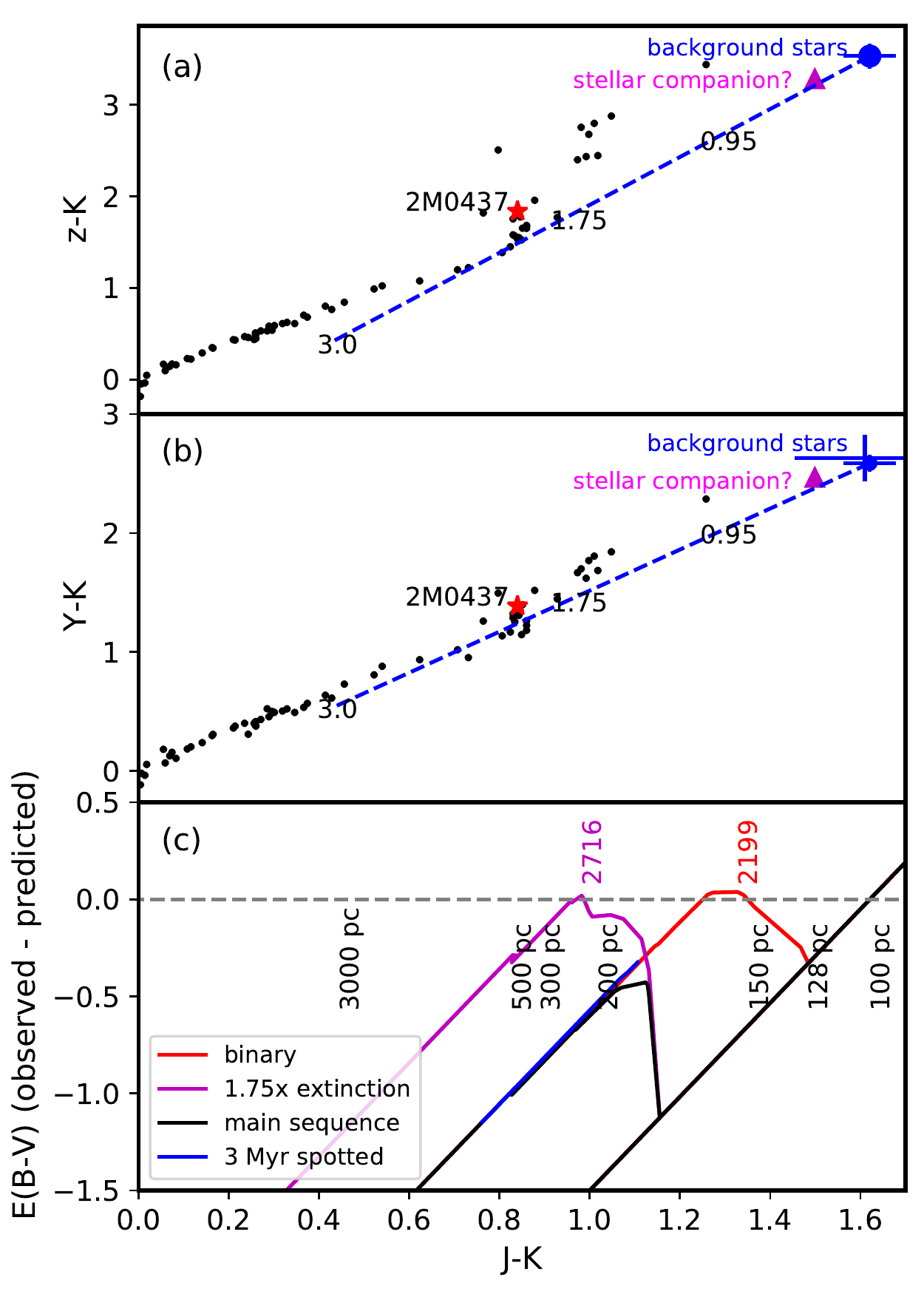}
    \caption{(a): $Z-K$ vs. $Y-K$ colors of 2M0437 (magenta star), the E background star (blue point), and photometric calibration stars (black points) from \citet{Hewett2006}.  The  co-moving stellar companion (magenta triangle) is presented in Sec. \ref{sec:co-moving}.   The blue dashed line is the predicted de-reddening trend using the extinction coefficients of \citet{Yuan2013} for $JHK$ and \citep{Gonzalez-Fernandez2018} for $ZY$.  The locations for key values of \ebv\ are indicated: 0.95 from the \citet{Green2018} map and 1.75 as a minimum value for consistency with an empirical dwarf star locus (see text).  (b):  Same as (a) but for $Y-K$ and including both the E and SW background stars (blue points).  (c) Difference between \ebv\ from the 3-d reddening map of \citet{Green2018} and the \ebv\ required for a star to be consistent with the \emph{observed} $J-K$ color and $K$ magnitude, as a function of the star's intrinsic $J-K$ color (and hence, given a color-magnitude relation, absolute $K$-magnitude and distance).  Selected corresponding distances, including that of 2M0437 (128 pc) are indicated.  The black, blue, and red, and magenta lines are the empirical main sequence of \citet{Pecaut2013}, the 3 Myr theoretical isochrone of heavily spotted (85\% spot coverage) stars of \citet{Somers2020}, unresolved equal-mass binaries, and the \citet{Pecaut2013} main sequence with an augmentation in reddening at all distances by a factor of 1.75.  Corresponding values of \teff\ where the last two cases achieve agreement between the predicted and observed \ebv\ are indicated.}
    \label{fig:colors}
\end{figure}

$ZYJK$ photometry is available for these stars from UKIDSS.  The $J-K$, $Y-K$, and (for the E star) $Z-K$ colors are consistent with the colors of highly reddened cool dwarfs (Figs. \ref{fig:colors}ab), however the required reddening (\ebv$>1.7$) is significantly more than that expected based on the \ebv=0.95 found in the reddening map of \citet{Green2018}.  Since these objects were not detected by \gaia\ and parallaxes are not available, a wide range of distance and luminosity combinations are possible.  We attempted to constrain the properties of these stars by establishing agreement between the \ebv\ at a given distance from a 3D reddening map based on independent observations of other stars \citep{Green2018}, and the \ebv\ needed for a star at that distance to explain both the observed $K$-magnitude and $J-K$ color.  Figure \ref{fig:colors}c plots the inconsistency (difference) vs. $J-K$ color, which is also correlated with distance.  Neither an empirical main sequence \citep{Pecaut2013} nor a 3-Myr isochrone of heavily spotted (85\%) stars \citep{Somers2020} provides a consistent scenario.\footnote{Young, pre-main sequence M dwarfs are more luminous, but also hotter and have bluer $J$-$K$ colors, thus stellar youth cannot alone resolve the discrepancy.}  The problem is that these objects have $J-K \approx 1.4$, much redder than is typical of M dwarfs ($\approx$0.85).

A possible solution is an unresolved binary of ultra-cool (\teff$\approx 2200$) dwarf stars within or immediately behind the Taurus cloud.  It seems unlikely that two such binaries would appear within 7" of a random fixed point, particularly if they are not affiliated with the cluster itself.  Alternatively, the reddening through Taurus behind the location of 2M0437 could be significantly greater than that estimated by \citet{Green2018}.  For the brighter E star a range of scenarios is possible, from an M7 dwarf at 200 pc and \ebv= 1.3, to a K0 star at 2600 pc and \ebv = 2.8, but consistency with $Z-K$ and $Y-K$ colors suggests a mid M dwarf (Fig. \ref{fig:colors}ab).  The fainter SW star could be as far away as 4.4 kpc, requiring an \ebv\ as high as 2.8 as well.  In lieu of a high \ebv, a non-ISM-like extinction law, i.e. larger grains and reduced wavelength dependence, could be invoked.  

The E star appears in many AO images and it can be used as a reference to estimate the brightness of 2M0437b in $H$-band and its $H$-$K$ color (Sec. \ref{sec:comp_photometry}).  $H$-band photometry is not available from UKIDSS; instead $H$-band photometry was performed on the E star by analyzing IRCS data with a short integration in which the signal from the unsaturated image of 2M0437 was measured, and a tandem, longer integration in which the signal from the E background star was measured, and then corrected for the ratio of integration times.  This yielded an $H$ magnitude for the E star of $17.50 \pm 0.02$.  We also use the simulations described above to perform an independent estimate of the reddened $H-K$ color and thus $H$ magnitude of the background stars.  We find $H-K = 0.47-0.63$, and in turn, the apparent H-band magnitudes of $17.23\pm 0.11$ and $18.34\pm 0.17$ for the E and SW stars, respectively;  where the uncertainties are taken to be the sum of the $K$-band error plus half the range of inferred $H-K$ values.  We adopt the average of the two estimates for the E star: $H=17.37 \pm 0.10$.

\subsection{Close-in Companion}
\label{sec:source}

\subsubsection{Astrometry}
\label{sec:comp_astrometry}

The close-in candidate companion was detected in 7 distinct epochs of AO imaging spanning 3 years (2 with Subaru/IRCS, 5 with Keck-II/NIRC2), enabling us to test whether it co-moves with 2M0437. Given the substantial absolute proper motion of 2M0437, the relative position of a fixed background star would have changed by $\mu_{rel} = 28$\,mas~yr$^{-1}$ or $\Delta \vec{x} = 85$\,mas over this interval, almost entirely as a $-5.5\degr$ change in position angle. Conversely, a bound companion in a circular face-on orbit ($a = 118$\,au) would exhibit orbital motion of $v_{orb} = 1.1$\,\ks, $\mu_{rel} = 2$\,mas~yr$^{-1}$, or $\Delta \vec{x} = 6$\,mas over this interval. The distribution of field star proper motions is a complicated function of brightness and galactic latitude/longitude, but a search of the Gaia EDR3 catalog \citep{Gaia2020} shows that the nearest other star that co-moves with 2M0437 to within $\mu_{rel} < 2$\,mas~yr$^{-1}$ has an angular separation of $\rho > 1000\arcsec$ (but see Sec. \ref{sec:co-moving}), and hence the empirical odds of an unassociated field star being seen in chance alignment appear to be small.

\begin{figure*}
	\includegraphics[width=2\columnwidth]{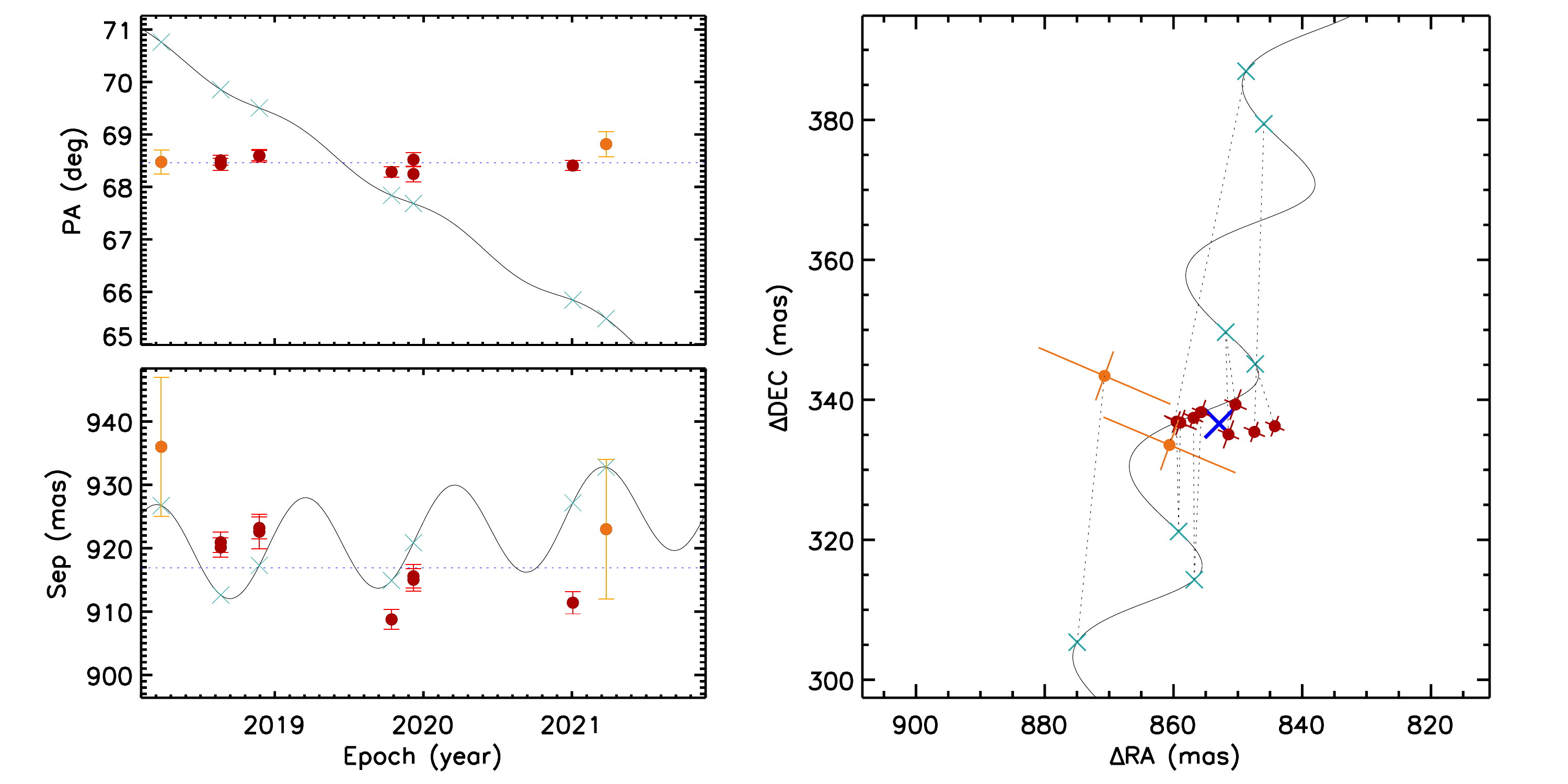}
    \caption{Astrometry of the companion 2M0437b.  Position angle and separation are plotted on the left side, while R.A. and declination relative to 2M0437 in mas are plotted on the light.  The solid curves are the predictions for a star at the distance of Taurus that is motionless in the \gaia\ reference frame.  The dashed blue lines are the predictions for a hypothetical co-moving object.  The predicted positions (for the observation epochs) of 2M0437b were it a background star are plotted as the blue crosses.  Red points are NIRC2 and orange points are IRCS data.  A 0.25 deg offset has been applied to combine these data (see text).}  
    \label{fig:astroB}
\end{figure*}

As we show in Figure~\ref{fig:astroB}, the relative position of the close-in companion remained nearly constant across the monitored interval, inconsistent with the trajectory expected for a background star ($\chi_{\nu}^2 = 97$ with 18 degrees of freedom), but much more consistent with zero ($\chi_{\nu}^2 = 4.47$ with 18 degrees of freedom). Furthermore, either the IRCS dataset or the NIRC2 dataset alone would support this conclusion at high confidence. We therefore conclude that the close-in companion is co-moving with 2M0437 and, at a projected separation of about 118 au, likely bound (see below),  justifying its designation as 2M0437~b. 

A linear fit to the relative astrometry yields a relative motion of $\mu_{rel,\rho} = -5.0 \pm 0.8$\,mas~yr$^{-1}$ and $\mu_{rel,PA} = -0.7 \pm 0.7$\,mas~yr$^{-1}$, with substantially improved goodness of fit over the purely co-moving case ($\chi_{\nu}^2 = 2.36$ with 16 degrees of freedom). The improvement in the quality of this fit raises the intriguing possibility that we are resolving orbital motion, and hence that we can add 2M0437~b to the small set of wide-orbit planetary-mass objects with resolved orbits (e.g., \citealt{Bryan2016,Pearce2019,Bowler2020}). Given the apparently edge-on geometry of the star's rotation (Section 3.1.5), a radial trajectory would provide strong circumstantial evidence of alignment between its orbit and the stellar spin. However, the inferred velocity ($v_{rel} = 2.6$ km/s) is substantially larger than the circular-orbit velocity ($v_{orb} = 1.1$ km/s if $a = 118$ AU) and could be explained with moderate inflation of the astrometric uncertainties, so this tentative result should be confirmed with additional monitoring in future seasons.

\subsubsection{Photometry}
\label{sec:comp_photometry}

Photometric analysis of NIRC2 images is complicated by the complex point-spread function of the Keck-2 telescope and its hexagonal segmented mirrors. We performed PSF fitting and subtraction of the primary star image for 7 independent observations at 4 epochs in $K'$-band.  The standard deviation (0.17 mags) is significantly larger than expected from the formal errors ($\chi^2=47.7$ for 6 d.o.f.), either due to systematics and/or intrinsic variability of the star and companion, equivalent to 0.17 mag uncorrelated extra error.  Adjusting for this the weighted mean contrast ratio is $\Delta K' = 6.82 \pm 0.07$ mag, for a $K_s$ magnitude of $17.21  \pm 0.07$.

Performance of the AO system degrades with shorter wavelength and $H$-band imaging is significantly inferior to that in the $K'$-band.  In our 25 November 2018 $H$-band imaging with NIRC2 both the companion and the "E" background star appear in the field.  Using the azimuthal median-subtracted image, we estimated the signal from the companion by a circular aperture centered at the location of the companion and subtracting the median of the counts in five identical apertures placed at locations rotated by 60 deg intervals around the center of the PSF.  The standard error is assumed to be dominated by background uncertainty and taken to be the standard deviation of the five divided by $\sqrt{5}$. The aperture is varied; the counts has a maximum of $100400 \pm 21000$ counts with a 700 pixel (0\arcsec.3) aperture and is insensitive to exact choice of apertures from 400-1000 pixels.  We determined the counts from the "E" background star using the same aperture and, subtracting the mean of the counts obtained in four aperture placed around the source aperture, obtained 244000 counts, arriving at a contrast of $0.96 \pm 0.2$ mags.  Combining this with the $H$-band estimate for E (Sec. \ref{sec:back_photometry}) we arrive at $H = 18.19 \pm 0.23$.  
We also performed a photometric analysis for the combined IRCS image in a similar fashion;
the photometric count was measured for the companion "b" using the median-subtracted image. 
To estimate the reduction of the photometric count and its systematic error due to the median subtraction, 
we implemented injection and recovery simulations placing a same-magnitude source at many different position angles, but with the same separation from the primary star. Based on these simulations, 
the $H$-band magnitude from the IRCS AO imaging was found to be $H = 18.03 \pm 0.09$.
We perform a weighted average of the two estimates for a final value of $18.05 \pm 0.08$.  Combining with the $K$-band magnitude we estimate the companion to have $H-K = 0.85 \pm 0.11$.

\subsubsection{Mass}
\label{sec:mass-age}

\begin{figure*}
    \includegraphics[width=0.49\textwidth]{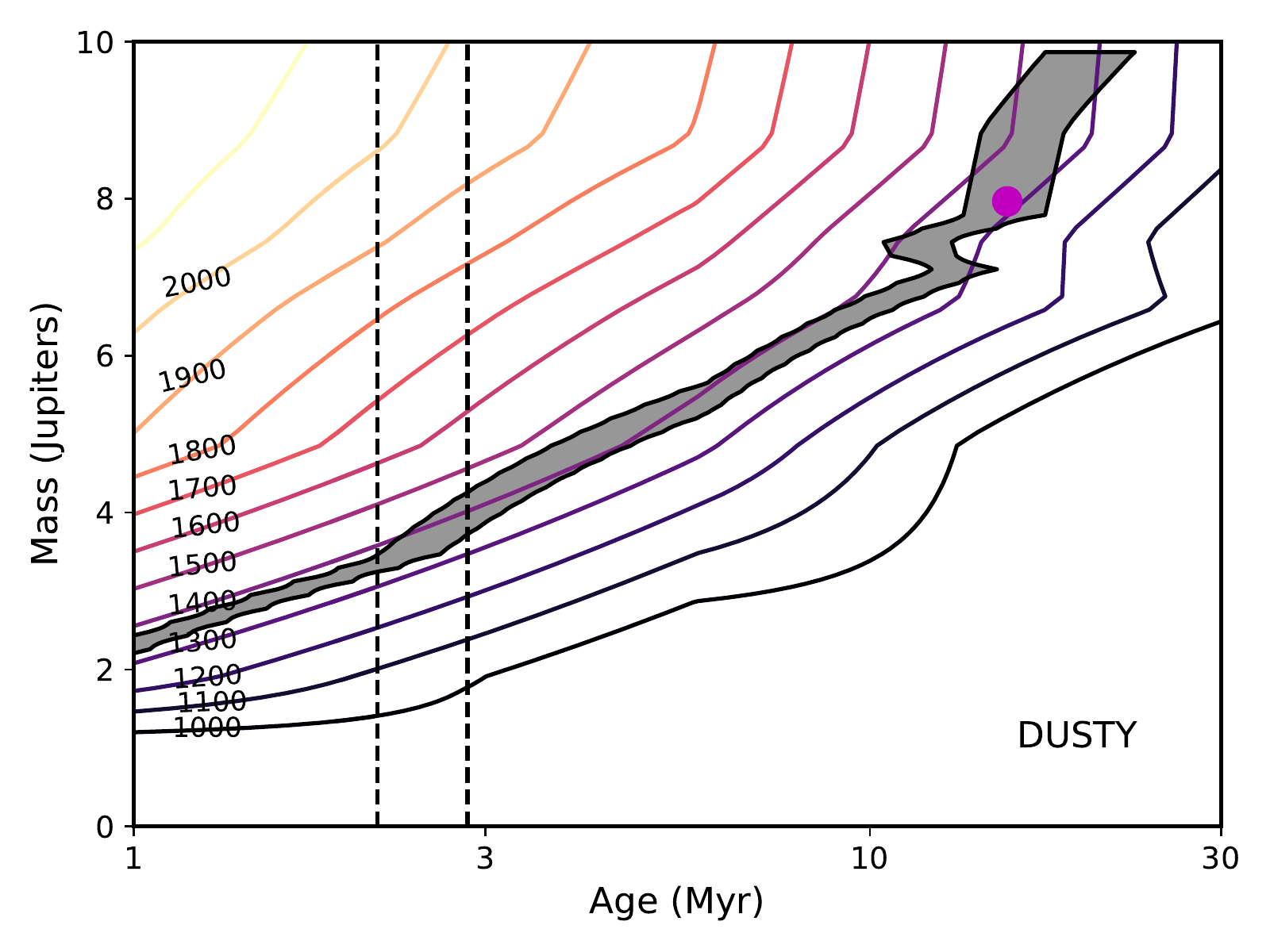}
     \includegraphics[width=0.49\textwidth]{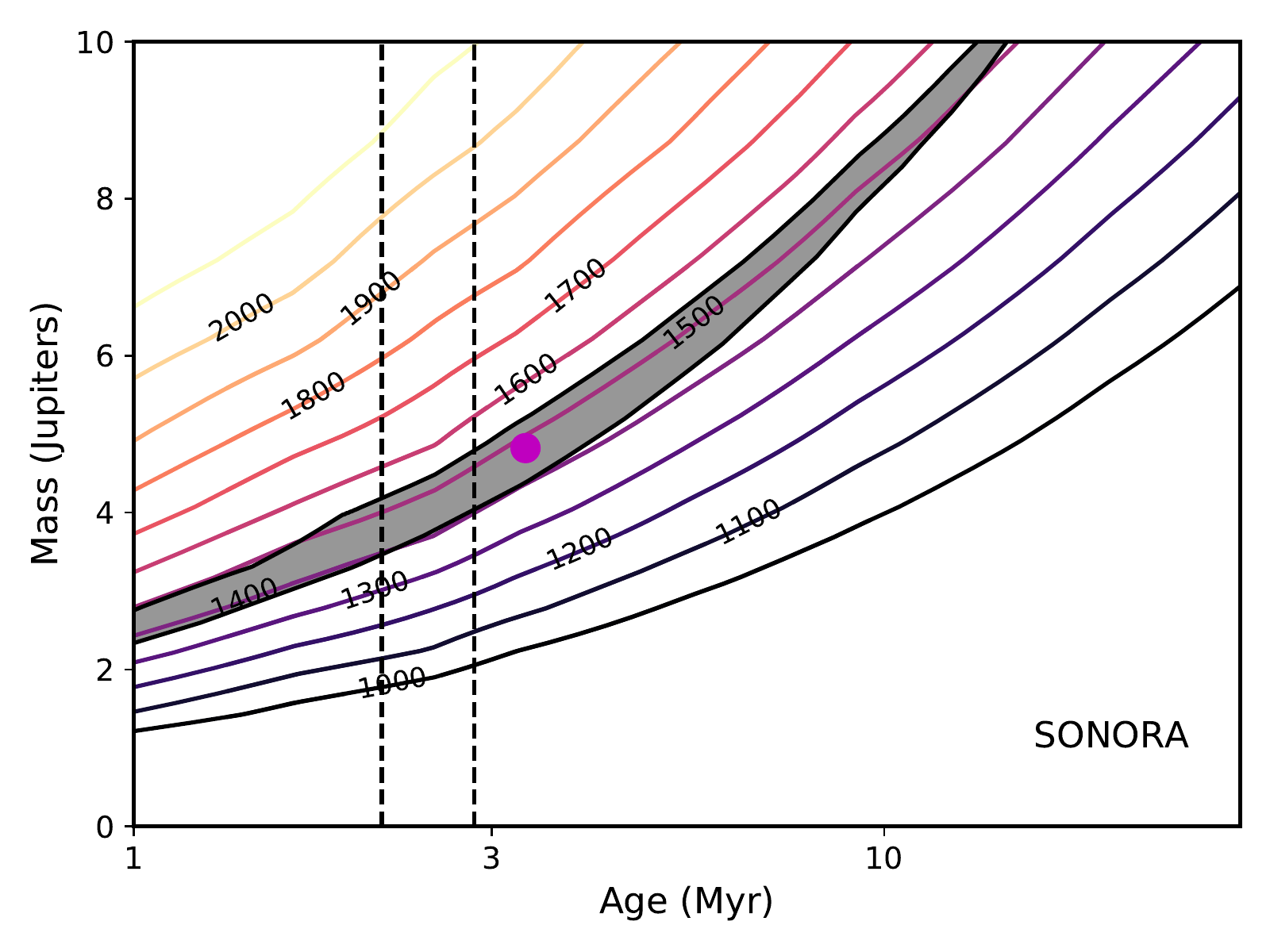}
    \caption{Model masses and ages that are consistent with observations of 2M0437b, i.e. $\Delta \chi^2 < 6$ between predicted and observed $K_s$ magnitude using the DUSTY models of \citet{Chabrier2000} and \citet{Baraffe2002} (left), and the SONORA models of M. Marley (in press) (right). For the SONORA models, we used an empirical relation between $K$-band bolometric correction and \teff\ \citep{Pecaut2013}.  Although this was relation was developed for older field stars, \citet{Filippazzo2015} showed that the $K$-band corrections are similar for young and field stars.   In each case, model effective temperatures are plotted as colored contour lines, the location of minimum $\chi^2$ as a magenta point, and the age range suggested by the association with the $\sim$2.5 Myr-old D4 Taurus subgroup \citep{Krolikowski2021} is demarcated by vertical dashed magenta lines.}
\label{fig:models}
\end{figure*}

To infer the mass of 2M0437b for a given age, we compare the object's absolute $K$ magnitude to two sets of models: the DUSTY models of low-mass stars and brown dwarfs with \teff=900-2800\,K of \citet{Chabrier2000} and \citet{Baraffe2002}, which include dust formation and opacity, and the cloud-free SONORA models of Marley et al., in press, for \teff=200-2400\,K.  All models assume solar metallicity and a solar C/O ratio.  We performed bi-linear interpolation of the published models on finer grids of mass and $\log$ age, and computed the $\chi^2$ statistic, and show zones of 95\% confidence ($\Delta \chi^2 < 6.0$, two degrees of freedom) vs age and mass in in Figure \ref{fig:models}.  We plot the locations of minimum $\chi^2$ in this parameter space (magenta points), but the degeneracy between mass and age means that wide ranges of the two parameters are allowed by the photometry.  The two sets of models largely give similar results, predicting a \teff\ of 1400-1500\,K and corresponding late L spectral type for 2M0437b.  Within the 2-3 Myr age range suggested by the association of 2M0437 with the D4 Taurus subgroup of \citet{Krolikowski2021} (vertical dashed magenta lines in Fig. \ref{fig:models}) the corresponding mass is 3-4\mjup.  

\subsection{A Second, Wide-separation, Co-moving Object}
\label{sec:co-moving}

\begin{figure}
	\includegraphics[width=\columnwidth]{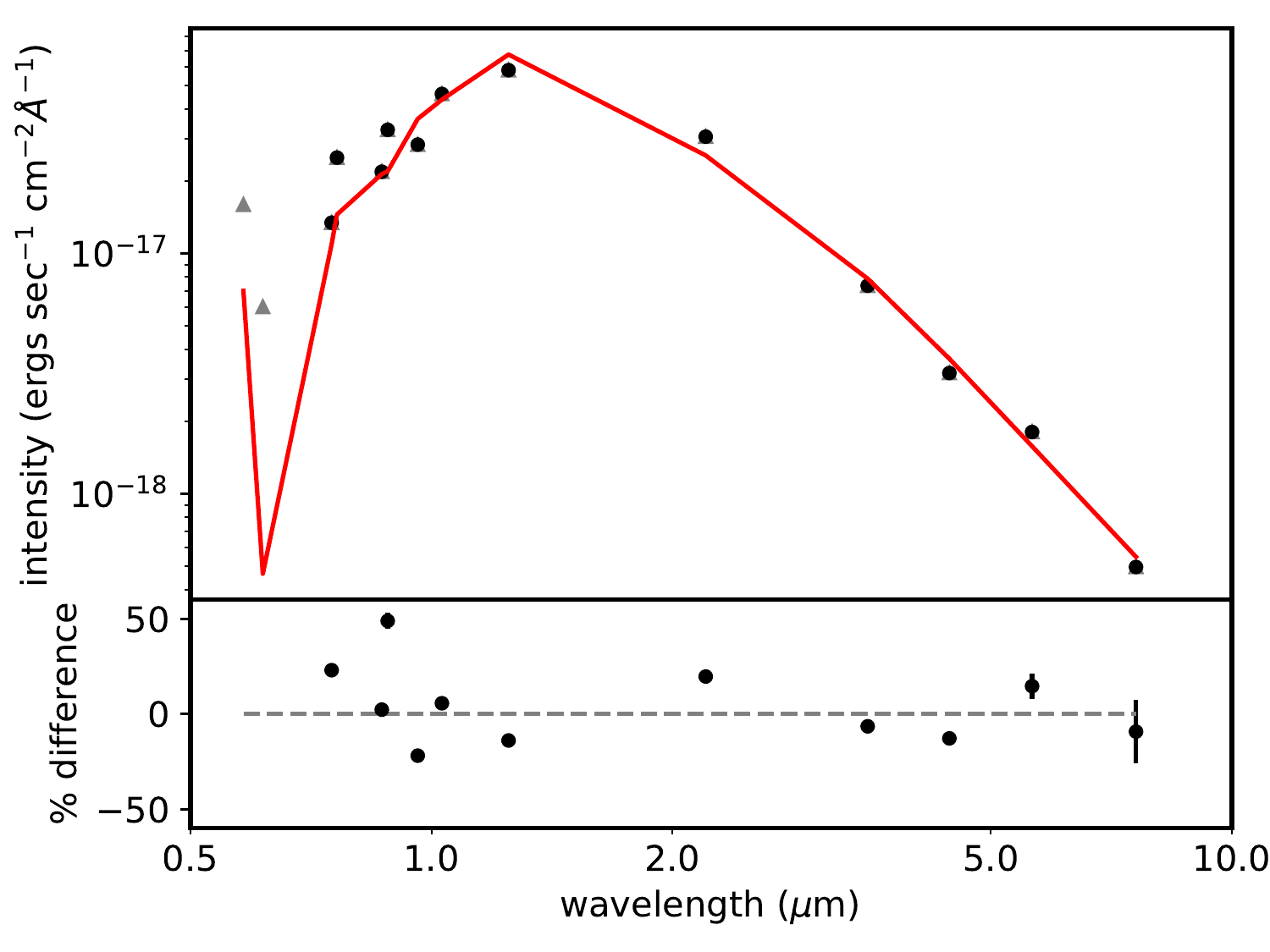}
    \caption{Spectral energy distribution of \comover, the wide separation co-moving companion to 2M0437, compared to the best fit AMES-Dusty model (red curve, \teff=2000\,K, log~g=3.5, [Fe/H]=0).  The $r$-band data (grey points) were excluded because of a possible significant contribution by \halpha.}  
    \label{fig:sed_c}
\end{figure}

\begin{figure}
	\includegraphics[width=\columnwidth]{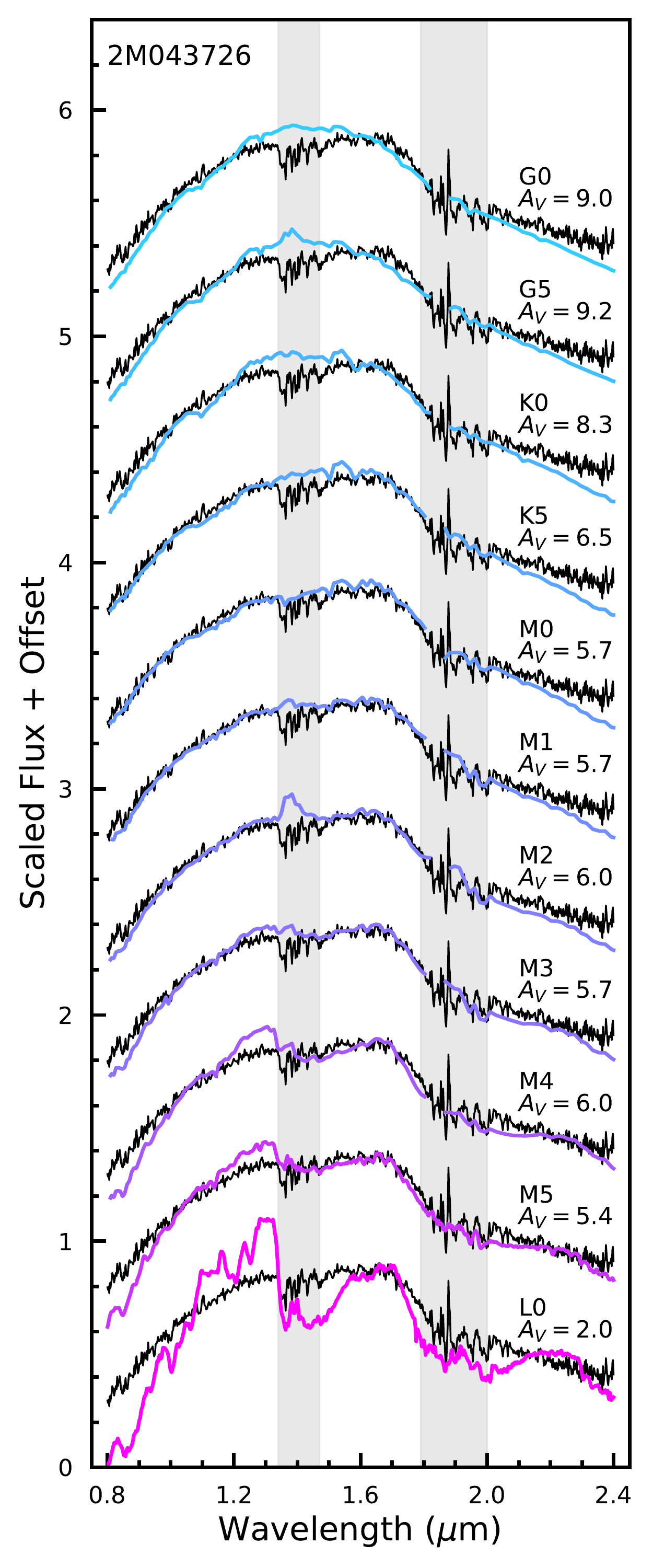}
    \caption{IRTF/SpeX prism spectrum of \comover\ (black curve) compared to fits of reddened template spectra of ultra-cool dwarfs \citep{Burgasser2004} and low-mass stars \citep{Cushing2005,Rayner2009}.  Grey zones mark wavelength ranges of telluric contamination which were excluded from the fitting.  The template spectra are of HD 109358 (G0), HD 165185 (G5), HD 145676 (K0), HD 36003 (K5), HD 19305 (M0), HD 42581 (M1), Gl 806 (M2), Gl 388 (M3), Gl 213 (M4), 2MASS J22120345+1641093 (M5), and 2MASS J0345432+254023 (L0). Reddening was calculated using the extinction law of \citet{Schlafly2011} and the extinction was found by minimizing $\chi^2$ with respect to the template spectrum down-graded in resolution to that of the observations ($\lambda/\Delta \lambda \approx 75$).} 
    \label{fig:spec_c}
\end{figure}

Using the \gaia\ EDR3 database \citep{Gaia2020}, we searched for additional objects that are co-moving with 2M0437 out to a radius of $\rho = 1000$\arcsec\, and identified a single, faint source (\gaia\ $G = 20.73$ mag) with a separation of 75\arcsec\ and with a proper motion that is only 2.05 mas~yr$^{-1}$ (0.92$\sigma$) discrepant from that of 2M0437.  The parallax ($8.3 \pm 2.7$ mas) is not very precise, but is indistinguishable from 2M0437 to within 0.18$\sigma$ and consistent with the Taurus population as a whole.  While a fit of unreddened DUSTY models to a SED constructed from the available photometry (Table \ref{tab:comoving}) suggests $\teff\approx$2000\,K (Fig. \ref{fig:sed_c}), if the extinction is allowed to vary, the range of possible \teff\ or spectral types is ranges from mid M- to G-type while the extinction ranges from 3 to 7 magnitudes in $V$-band.  Our low-resolution $JHK$ spectrum of the object (Sec. \ref{sec:modresspec}) lacks any of the deep molecular features associated with ultra-cool dwarfs but is consistent with spectra of extremely reddened ($A_V \gtrsim 5$) low-mass stars (Fig. \ref{fig:spec_c}).  The best fits with a standard reddening law is to a M1 spectral type ($\chi^2=1948$, 661 degrees of freedom), with the 95\% confidence interval ($\chi^2 < 3243$) spanning spectral types K5-M3.   We provisionally refer to this object as \comover, to distinguish it from 2M0437 but avoid a potentially ambiguous letter designation.  

\comover\ could either be a binary companion to 2M0437, or merely a fellow member of the D4 group.  Its parallax is not sufficiently precise to discriminate between these possibilities, and, at 10,000 AU, the escape speed is a mere 180\,m~sec$^{-1}$, well within the error of \gaia\ astrometry.  The probability of a chance association of another, unbound Taurus member is not negligible, but difficult to quantitatively assess due to the highly correlated distribution of cluster members on the sky.  Statistically, the unassociated member scenario is only likely where the surface density on the sky approaches $>100$~pc$^{-2}$ \citep[see Table 1 in ][]{Joncour2018}, but a more extensive search of the \gaia\ EDR3 database around the location of 2M0437 shows that in this region of Taurus the density is $\la$1~pc$^{2}$ and the nearest other Taurus member (2MASS J04353164+2715081) is at a separation of 34\arcmin (1.3 pc).   Based on the two-point correlation function of Taurus members, \citet{Kraus2008} found that bound binaries and unbound neighbors are on average equally common at a projected separation of $\rho \sim$17,000\,AU, whereas bound companions are approximately twice as likely at $\rho = $9,600 AU.   Assuming that the semi-major axis is approximately the observed separation, the orbital period of the system will be of order a Myr.

\begin{figure}
\centering
 \includegraphics[width=\columnwidth]{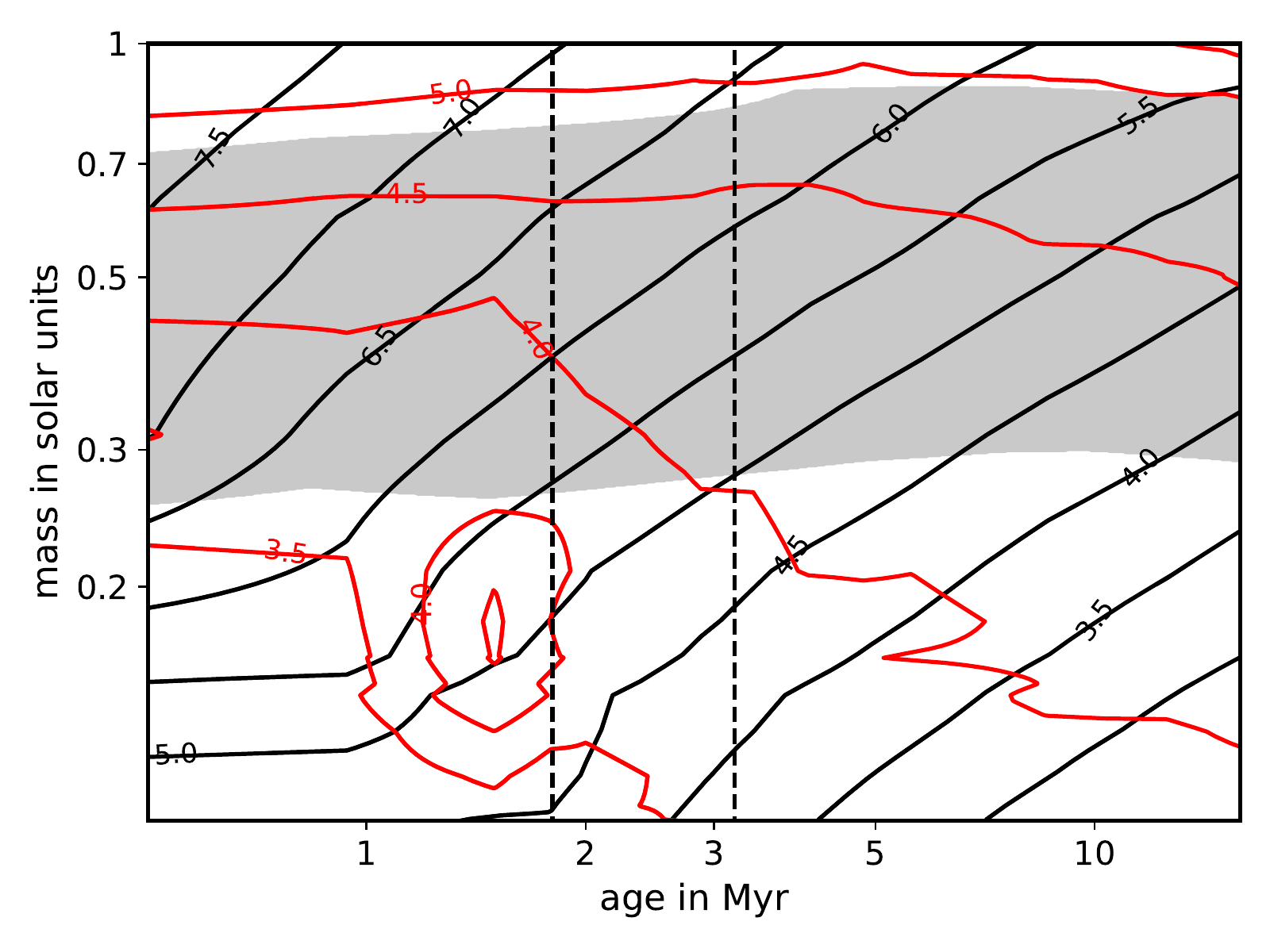}
\caption{The combinations of standard (ISM-like) extinction (red contours of $A_V$) and ``grey", wavelength-independent extinction (black contours) that can explain the apparent brightness of \comover, the co-moving stellar companion to 2M0437, as a function of age and mass.  The theoretical pre-main sequence models of \citet{Baraffe2015} and the reddening coefficients of \citet{Yuan2013} and \citet{Gonzalez-Fernandez2018} are adopted, and the more precise parallax of 2M0437 is used.  The grey zone represents the constraint from an infrared spectrum (Fig. \ref{fig:spec_c}), which limits the spectral type to M3-K5, corresponding to \teff\ of 3360-4140\,K for pre-main sequence stars \citep[][grey zone]{Pecaut2013}.  The vertical dashed line is the estimate age range for the D4 Taurus sub-cluster \citep{Krolikowski2021} with which \comover\ and 2M0437 are kinematically affiliated.}
\label{fig:extinction}
\end{figure}

The required extinction ($A_V \gtrsim 5$ for a standard extinction law) to explain the spectrum and photometry of \comover\ exceeds estimates of the \emph{total} extinction through Taurus cloud in this region ($A_v \approx 3$, see Sec. \ref{sec:back_photometry}) and is in marked contrast to 2M0437, which is virtually unreddened.  While the colors of \comover\ also indicate that it is extincted by several magnitudes (Fig. \ref{fig:colors}), consistent with that required to explain its $JHK$ spectrum, it is still too faint in the near-infrared for this to be following a standard extinction law alone.  We can approximate the appearance of \comover\ by two-component extinction; one with small dust grains causing standard extinction, and the other with large grains causing ``grey" (wavelength-independent) extinction.  The latter must have a grain size $\gg$10 $\mu$m, since the SED closely follows the Rayleigh-Jeans law through $\lambda=7.6\mu$m (Fig. \ref{fig:sed_c}).   We computed the magnitudes of these components by comparing the $zK$-band photometry of \comover\ to the pre-main sequence models of \citet{Baraffe2015}, assuming the star has the same (more precise) parallax as 2M0437.  $Z-K$ color agreement constrains standard extinction and the residual difference between the observed and predicted $K$-band magnitude sets the grey extinction (Fig. \ref{fig:extinction}).  Based on the range of spectral types (K5-M3) consistent with $\chi^2$ fits of reddened template spectra to our $JHK$ spectrum (Fig. \ref{fig:spec_c}), and a relation between spectral type and \teff\ for pre-main sequence stars \citep{Pecaut2013}, we constrain \teff\ to 3360-4240 K (grey area in Fig. \ref{fig:extinction}).  The mass of \comover\ is then 0.25-0.8\msun, which means that it, not 2M0437, is likely the primary in the system!  If the age is $\sim$2.5\,Myr, then there is at least 4 mag of standard extinction and at least 5 mag of "grey" extinction along line of the sight.

A plausible explanation for the high level and two-component nature of extinction is that we are observing \comover\ through an edge-on circumstellar disk, with an outer part causing standard extinction and an inner part, where grain growth has proceeded,  responsible for grey extinction.  The anomalous low luminosity of several ultra-cool objects have been explained in this way.   \citep{Mohanty2007,Looper2010,Haffert2020,Christiaens2021}.  An alternative explanation is dust clouds in its atmosphere \citep[e.g.,][]{Skemer2011}, but this would be accompanied by extreme reddening, something not observed in this case (Fig. \ref{fig:color-mag}).  We also interpret the anomalous brightness of \comover\ in \gaia\ $G$- and Pan-STARRS $r$-bands (Fig. \ref{fig:sed_c}) as a result of elevated H$\alpha$ emission, and thus accretion.  Unfortunately, photometry at mid-infrared wavelengths that could detect such a disk is lacking: \comover\ was not detected in either the W3 (12-$\mu$m) or W4 (12-$\mu$m) channels of \wise\ and a \spitzer\ MIPS 24-$\mu$m observation just missed this object.   

\subsection{Limits on Additional Companions}
\label{sec:limits}

Multi-planet systems are common among the short-period planets discovered via transits and RVs, and an increasing number of directly-imaged planetary-mass companions are also being found to also have additional planetary-mass siblings. The canonical planetary system HR 8799 hosts at least four giant planets \citep{Marois2010}, and Beta Pictoris was also recently recognized to host a second giant planet \citep{Lagrange2019}. Among wider-orbit planets, the TYC 8998-760-1 system hosts two companions with $M \sim 5$--15 $M_{Jup}$ at projected separations of 160 au and 320 au \citep{Bohn2020}. The 2M0437 system possibly joins this short list, hosting a close ($\rho \sim 118$ au) companion, which motivated a robust determination of the observational limits on yet more companions.

\begin{figure}
	\includegraphics[width=\columnwidth]{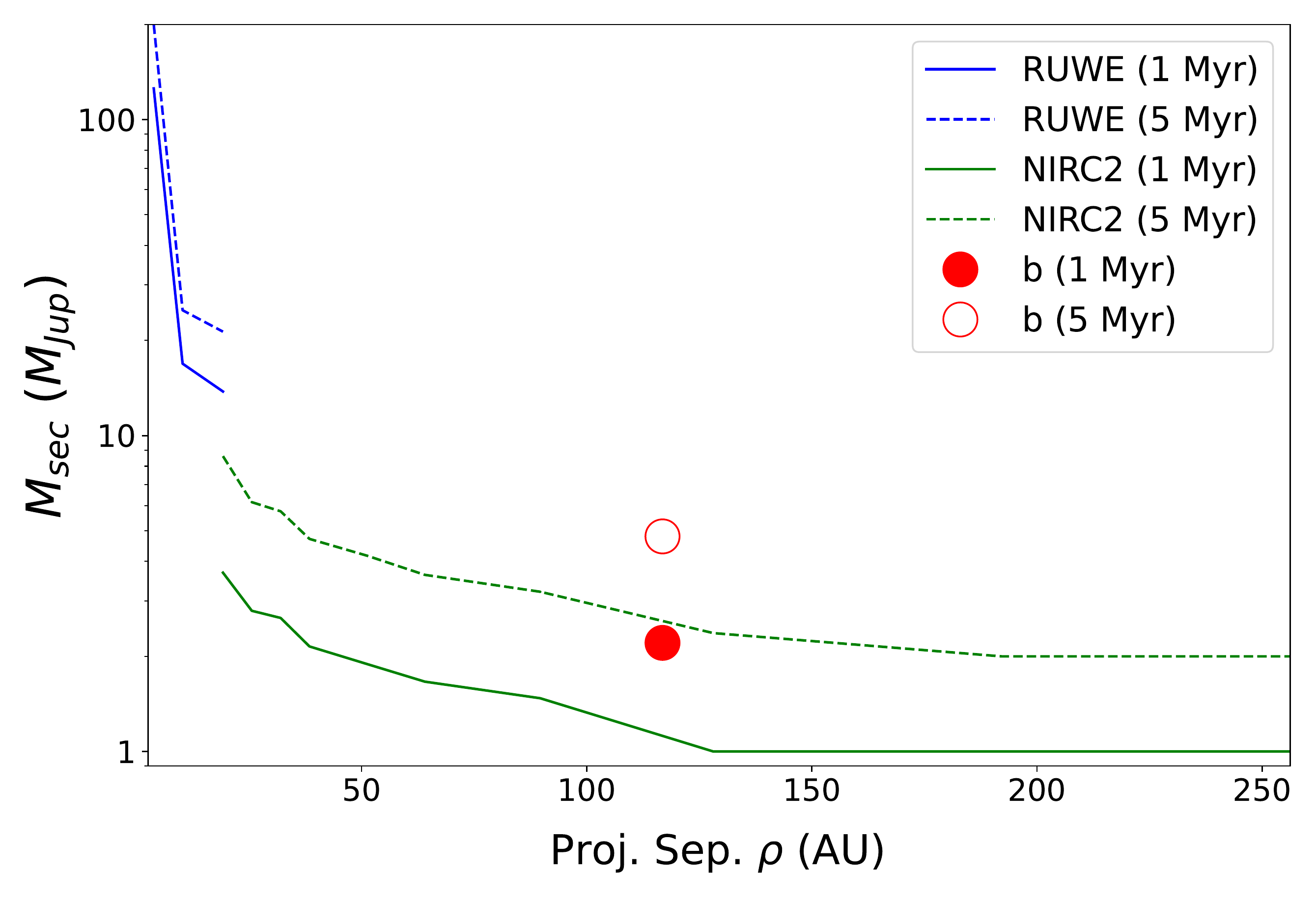}
    \caption{Detection limits for additional companions in the 2M0437 system, as derived from the AO observations with NIRC2 (green), or based on the absence of excess Gaia noise (blue), for assumed companion ages of 1 Myr (solid) and 5 Myr (dashed).}  
    \label{fig:detlims}
\end{figure}

In Table \ref{tab:ImgLim}, we present the observational detection limits, measured as contrast $\Delta m$ as a function of angular separation $\rho$, that we have derived for the two individual epochs that were deepest in $K'$. We found that an observation from 22 August 2018 was more sensitive at small separations (due to superior AO performance and the availability of well-matched PSF calibrators) and an observation from 9 December 2019 was more sensitive at wide separations (due to a longer total integration time). In Figure~\ref{fig:detlims}, we summarize the corresponding detection limits in terms of companion mass $M_{Jup}$ and projected physical separation in au, adopting the hot-start DUSTY models of \citet{Chabrier2000} and assuming companion ages of either 1 Myr or 5 Myr. We find that the contrast limits saturate to the lowest mass defined for the models ($1 M_{Jup}$ at 1 Myr or $2 M_{Jup}$ at 5 Myr) at separations of $\rho \ga 1 \arcsec$, and the limits remain below 10 $M_{Jup}$ to the innermost radius at which robust detection limits could be assessed ($\rho \ga 10$ AU).

The primary star was generally saturated in most imaging observations that were taken, but even if it was not, a robust calibration of the AO PSF at 1--3 $\lambda /D$ ($\rho \la 150$ mas) is typically quite challenging. We therefore instead place limits on close-in companions based on the absence of excess noise in the Gaia EDR3 observations \citep{Gaia2020}, as denoted by its Renormalized Unit Weight Error ($RUWE$; \citealt{Lindegren2018}). The measured value of $RUWE = 1.214$ would only be modestly elevated for photometrically stable field dwarfs (e.g., \citealt{Bryson2020,Wood2021}; Kraus et al., in prep) and is consistent with single-star values for photometrically variable young stars (S. Fitton \& B. Tofflemire, priv. comm.). Based on a calibration of the companion parameter space that would induce excess noise (\citealt{Wood2021}; Kraus et al., in prep), this corresponds to contrast limits of $\Delta G \sim 0$ mag at $\rho = 30$ mas, $\Delta G \sim 4$ mag at $\rho = 80$ mas, and $\Delta G \sim 4.5$ mag at $\rho = 150$ mas. In Figure~\ref{fig:detlims}, we also summarize the limits in terms of mass and projected physical separation, adopting the stellar evolutionary models of \citet{Baraffe2015} for either 1 Myr or 5 Myr. 

Finally, while the outer working angle of the NIRC2 observations was typically $\rho \sim 4$--5$\arcsec$, we can place a robust limit at wider separations based on the nonexistence of other comoving companions in the Gaia EDR3 catalog \citep{Gaia2020}. Aside from \comover, there are no sources within $\rho < 300\arcsec$ ($\rho < $40,000 AU) with proper motions and parallaxes that are consistent with 2M0437 to within 5$\sigma$. The nominal brightness limit of Gaia is $G_{lim} < 21$ mag, though in the vicinity of 2M0437, only sources brighter than $G < 20.6$ mag are consistently given five-parameter astrometric solutions. According to the evolutionary models of \citet{Baraffe2015}, the brightness limit of $G = 20.6$ mag is 1 magnitude deeper than the lowest mass defined at 1 Myr (10 $M_{Jup}$), but then would correspond to limits of 12 $M_{Jup}$ at 2.5 Myr or 15 $M_{Jup}$ at 5 Myr.

In summary, these limits demonstrate that there are no other stellar or brown dwarf companions in the 2M0437 system unless they are located very close to the primary star, with a limit of $\rho < 20$ AU for companions at the deuterium-burning limit and $\rho < 5$--6 AU for companions at the hydrogen-burning limit.


\section{Discussion and Summary}
\label{sec:discussion}

\begin{figure}
    \includegraphics[width=\columnwidth]{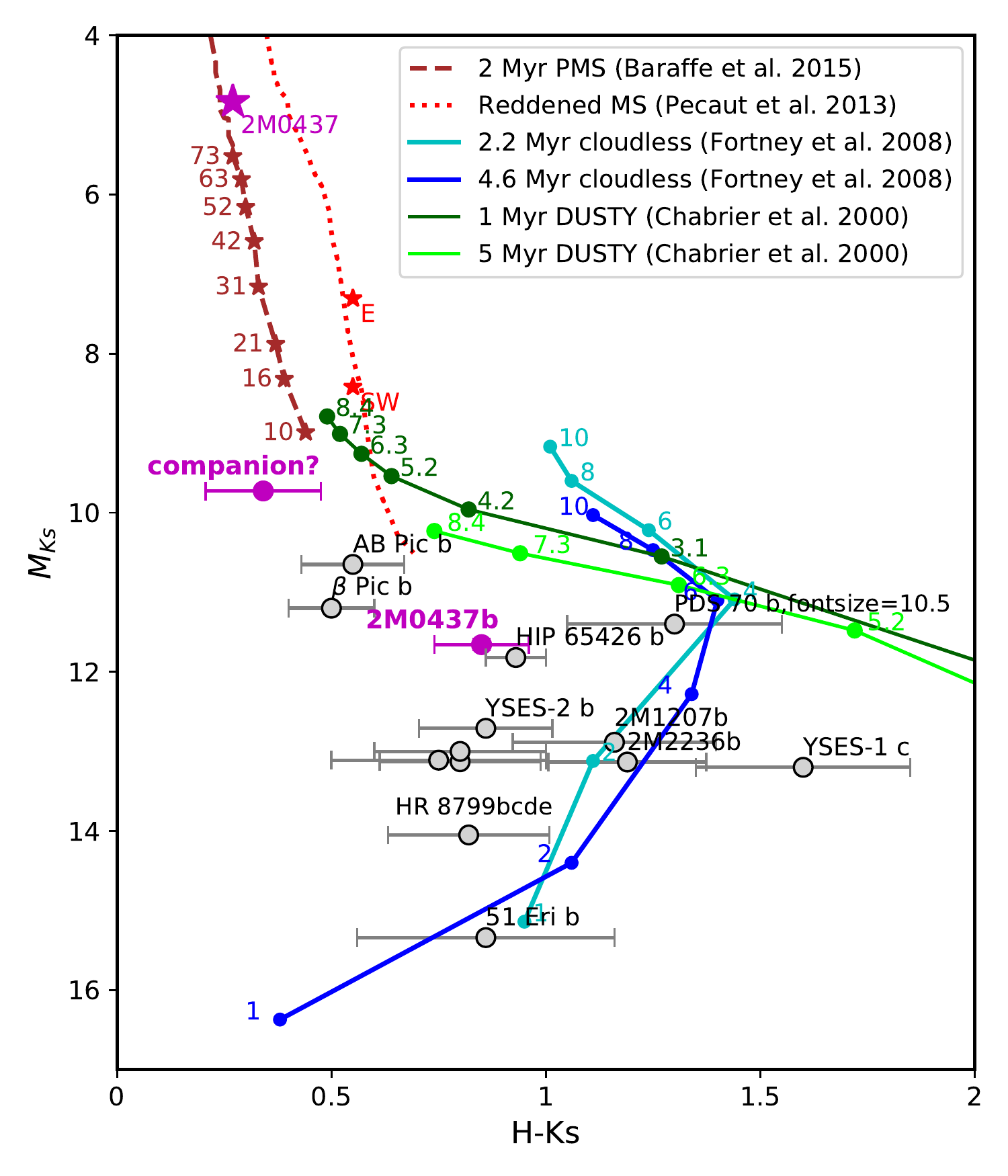}
    \caption{The absolute $K_s$ magnitude and $H$-$K_s$ color of 2M0437b (magenta circle), the companion to a Taurus M dwarf primary (magenta star), are similar to those of companions to young stars with inferred masses below the D-burning limit of 13\mjup\ (open black points  \citep{Marois2008,Marley2012,Bowler2017,Ward-Duong2021,Bohn2021}, and consistent with a model of 1~Myr-old cloudy Jupiter-mass planets \citep{Fortney2008} (cyan line with mass-annotated points).  They are inconsistent with those of pre-main sequence stars \citep[magenta isochrone,][]{Baraffe2015} or an empirical main sequence\citep{Pecaut2013} that has been significantly reddened (\ebv= 1.63, see text).  The red stars represent two sources that are 6\arcsec\ from 2M0437; here they are assumed to be 750 pc in the background but their actual distances are not known.}
\label{fig:color-mag}
\end{figure}

2M0437b falls amongst a sequence of moderately red ($H-K_s \sim 1$) objects in a color-magnitude diagram (CMD) that includes the planets of the HR 8799 system as well as several other directly imaged planetary-mass companions (Fig.~\ref{fig:color-mag}).  Like the bcde planets of the HR 8799 system, it is consistent with the colors of an $\approx$1400\,K dusty late L-type dwarfs \citep{Bonnefoy2016}.  In this CMD, 2M0437b most closely resembles HIP 65426b, a $\sim$8 \mjup\-mass object with \teff$\approx$1600 orbiting a star in the $\approx$10-20 Myr-old Centaurus-Crux association \citep{Chauvin2017,Cheetham2019}.  2M0437b is much younger, and therefore likely to be significantly less massive than HIP 65426b.  YSES-2b is a similar planet orbiting a member of the same association; it is less luminous than HIP 65426b, and thus predicted to be less massive \citep[$\sim$6\mjup,][]{Bohn2021}.  Other comparable objects include the four known planets (bcde) of the HR 8799 system.  These have estimated masses of 5, 7, 7 \citep{Marois2008} and 8 \mjup\ \citep{Currie2011}, respectively, comparable to 2M0437b but this system is significantly older \citep[30-60 Myr;][]{Marois2008,Zuckerman2011}, and thus these planets are much less luminous (Fig.~\ref{fig:color-mag}).  

Two sets of published models with reported $H-K$ colors are also plotted in this CMD (Fig. \ref{fig:color-mag}):  the cloud-free ``hot start" models of \citet{Fortney2008} and the ``dusty" models of \citet{Baraffe2002}.  Neither set of models explain the properties of 2M0437b and indeed, most known directly-imaged planets (Fig.\ref{fig:color-mag}).  (See \citet{Marley2012,Bonnefoy2016} for a detailed investigation of the HR 8799 system.)  One reason for this failure is that $H-K$ reddening by dust in the planets is sensitive to the poorly understood microphysical properties of the grains and settling in the convective atmospheres of these objects.  But based only on predicted absolute $K$ magnitude, the cloud-free models of \citet{Burrows1997} suggest a mass of 1-5\mjup, and the closest models of \citet{Fortney2008} have masses of 3-5\mjup, depending on the exact age, similar to the results from the analyses of the DUSTY and SONORA models (see Sec. \ref{sec:mass-age}).  Here, we have assumed that 2M0437b formed at the same time as its star: for a stellar age of $\sim$2.5 Myr this is not necessarily true, since the preferred "core accretion" model of giant planet formation typically require Myr (see below).  2M0437b could be even younger, and thus less massive than estimated here.  Further, the inferred mass depends sensitively on whether a high-entropy (``hot start") or lower entropy (``cold star") formation is assumed.  

\begin{figure}
\centering
 \includegraphics[width=\columnwidth]{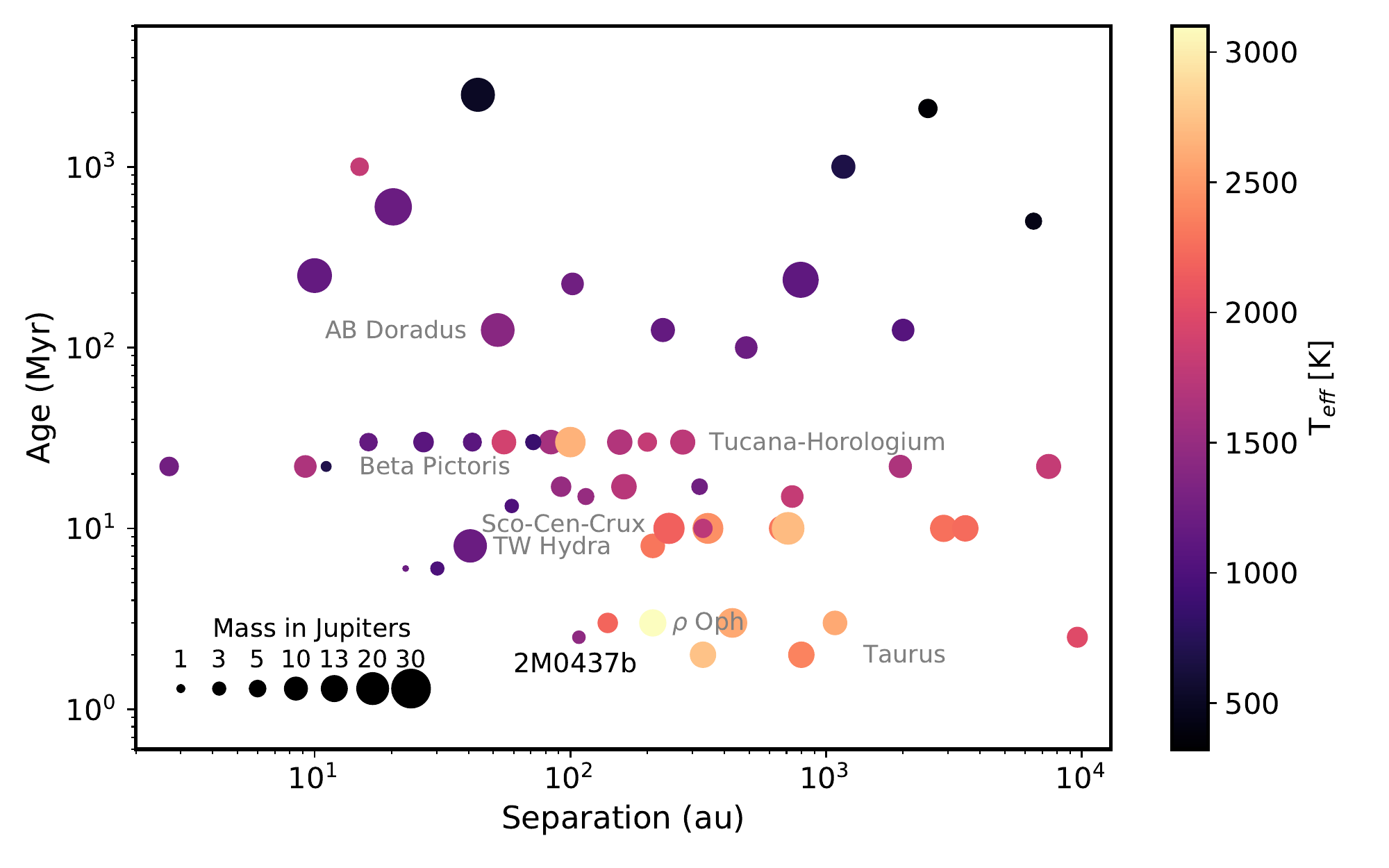}
\caption{2M0437b stands out among planet-mass companions detected by imaging in terms of its young age, proximity to the star, and low inferred mass. It is closer to the planet population expected to have formed within a disk.  The point size is scaled to object mass and color-coded by object effective temperature.  The young stellar associations in which these objects are located are labeled.} \label{fig:imaged}
\end{figure}

The number of confirmed substellar-mass companions has steadily increased since the beginning of the proliferation of AO imaging instruments two decades ago.  However, there remain few known exceptionally young objects, i.e., around stars of a few Myr age when disks are prevalent (Fig. \ref{fig:imaged}). With one possible exception, all of these have masses near or larger than the D-burning limit ($\approx$13\mjup).  These include two other members of Taurus: DH Tau B \citep{Itoh2005} has an inferred mass that is $\sim$15\mjup\ \citep{Bonnefoy2014}, i.e. near the deuterium-burning limit and its own disk \citep{vanHolstein2021}.  FU Tau b, the secondary of a brown-dwarf binary, has a similar inferred mass \citep{Luhman2009}.  (A yet lower mass candidate companion to DH Tau B, possibly in the Jupiter-mass range, has been reported by \citealt{Lazzoni2020}.)  CT Cha b, a companion to a member of the 1-5 Myr Chamaeleontis cluster \citep{Schmidt2008}, has an inferred mass in the brown dwarf regime \citep{Wu2015}.  ROXs 12 b, ROXs 42 Bb, and SR 12ABc all orbit members of the $\sim$3 Myr-old $\rho$ Ophiuchus star-forming region; the least massive of these is ROXs 42 Bb \citep{Currie2014,Kraus2014b}, with a mass of $10 \pm 4$\mjup \citep{Kraus2014b}.  

Only four\footnote{A possible fifth planet, GJ 504b \citep{Kuzuhara2013}, could be Jovian mass for a stellar rotation/activity-based age of $\sim$20 Myr, but is in the brown dwarf range if an isochrone age of $\sim$4 Gyr is correct \citep{Bonnefoy2018}.} known directly imaged planets have nominal masses comparable to or lower than 2M0437b ($\lesssim5$\mjup), and all are older:  PDS 70 b and c, which orbit a $\gtrsim5$ Myr-old star in the Scorpius-Centaurus region \citep{Keppler2018,Haffert2019}, 51 Eri b, which orbits a member of the $\sim$23 Myr-old Beta Pictoris Moving Group \citep{Macintosh2015} and HD 95086b \citep{Rameau2013}, with a host star that has been recently re-assigned to the $\sim$13 Myr-old Carina association \citep{Booth2021}. Notably, all the host stars are much more massive than 2M0437.

2M0437 lacks any indication of an infrared excess due to an accretion disk, although we cannot rule out a disk that has been cleared inside $\approx$2 au, or a debris disk.  \citet{Luhman2009} found that about half of all low-mass stars in Taurus lacked disks and inferred a timescale for the clearing of inner disks to be $\sim$0.5 Myr, substantially shorter than the age of 2M0437 of $\sim$2.5 Myr.  While the "disk-free" stars have been proposed to include an older ($\gtrsim$10 Myr) component of Taurus \citep[e.g.,][]{Kraus2017}, and 2M0437 seems to be most consistent with one of the more distributed sub-populations of Taurus \citep{Krolikowski2021}, the age of that sub-population is consistent with the age of many stars in the core populations of Taurus. The kinematic relation of those older stars to Taurus is debated \citep{Luhman2018,LiuJ2021}, and, regardless, 2M0437 is certainly not among them.

The existence of the super-Jupiter 2M0437b at an age as young as $\sim$2 Myr poses a challenge to both of the leading scenarios for giant planet formation: core accretion and disk instability  \citep{Dangelo2018}.  Core accretion requires time for the formation of a sufficiently massive core ($\sim5M_{\oplus}$) and for accretion of disk gas to the point of the runaway state.  These timescales grow with separation $a$ as $a^{3/2}M_*^{-1/2}$ from a central star of mass $M_*$, and thus this is not considered a viable means to produce planets at large separations around very low-mass stars.  On the other hand, gravitational instability in disks is very rapid, but requires that the surface density $\Sigma$ of the disk exceed a value set by the Toomre condition $\Sigma > c \Omega/(\pi G)$, where $c$ is the gas sound speed, $\Omega$ the angular orbital speed, and $G$ the gravitational constant.  This condition places a requirement on disk mass ratio $\sim c\Omega a^2/(GM_*) = c/v_K$, where $v_K$ is the Keplerian speed, which becomes problematic for low $M_*$ and high $a$.  While $c$ cannot be less than a few hundred m~sec$^{-1}$, $v_K$ falls to $\sim$1 km~sec$^{-1}$ at 100\,au from 2M0437, meaning that disk mass fractions approaching unity are required, contrary to observations that suggest mass fractions $\ll 1$ for very low mass stars \citep{Pascucci2016,Sanchis2020}.  Indeed, the inferred mass ratio of "b" (0.015-0.03) suggests that, if it formed from a disk around 2M0437,a large fraction was incorporated into the planet!  

If 2M0437 and \comover\ are coeval then this poses the question of why the disk of the former has completely dissipated while the latter appears to be still vigorously accreting.  One possible difference is mass; but the observed trend, that disks around more massive stars dissipate faster \citep{Ribas2015}, cannot be the explanation if the companion is more massive.  There is also a tendency for stars without detectable disks to rotate faster than those with disks \citep{Rebull2020} and 2M0437 follows this pattern (Fig. \ref{fig:periods}).  This is observed in other clusters, particularly among the late-type members, and is generally ascribed to a binary's more rapid dissipation of a disk which would otherwise drain angular momentum from the contracting pre-main sequence star \citep{Bans2012,Harris2012,Barenfeld2019}.  2M0437 is a single star, although the potential influence of a massive planet on a wide orbit on disk evolution such as 2M0437b has not been well explored.

2M0437b is a challenging target for conventional AO imaging that performs wavefront sensing in the optical where the host star is faint ($r = 15.1$).  Characteristic of the Taurus field with its high extinction, there are no brighter stars within a typical isoplanatic patch ($\sim$30\arcsec).  Where available, a laser guide star such as that at the Keck telescopes can be used, but low Strehl limits observations at shorter wavelengths.  A new generation of AO systems \citep[e.g.,][]{Bond2018,Wizinowich2020} will permit high Strehl imaging of fainter host stars at $\lambda >2$\micron.   On the other hand, the lack of any substantial disk means that direct imaging observations are free of the confounding or obfuscating material material that have been encountered in other systems \citep[e.g.,][]{Currie2019,Stolker2020}.  Long-term astrometric monitoring of "b" could reveal its orbit and better constrain the host star's mass and age.  Both objects are appropriate targets for observations by the {\it Hubble Space Telescope}, i.e. multi-band photometry with Wide Field Camera 3, to constrain \teff\ and gravity, and search for molecular absorption features in their atmosphere.  

\begin{table*}=
\caption{Summary of AO Observations for Astrometry \label{tab:TabNIRC2}}
\begin{tabular}{lrcrrrccr}
\hline
Name & Epoch & Filter & $N$ & $\rho$ & P.A. & $\Delta m$ \\
& (MJD) & & & (mas) & (deg) & (mag) \\
\hline
\multicolumn{7}{c}{Subaru + IRCS}\\
\hline
b & 58206.231 &      $H$ &   10 & 936$\pm$11 &  68.48$\pm$0.23 &  ... \\
b & 59300.231 &      $K_c$ &   5 & 923$\pm$11 &  68.82$\pm$0.24 &  ... \\
SW & 58206.231 &      $H$ &   10 & 6249$\pm$12 & 238.925$\pm$0.018 &  ... \\
SW & 59300.231 &      $K^\prime$ &   5 & 6236$\pm$30 & 239.62$\pm$0.11 &  ... \\
E & 58206.231 &      $H$ &   10 & 6731$\pm$18 &  79.168$\pm$0.020 &  ... \\
\hline
\multicolumn{7}{c}{Keck-2 + NIRC2}\\
\hline
b & 58352.598 &     $K^\prime$ &   4 & 920.1 $\pm$ 1.5 &  68.43$\pm$0.12 &  6.76$\pm$0.03 \\
b & 58352.602 &     $K^\prime$ &   4 & 920.9 $\pm$ 1.6 &  68.51$\pm$0.10 &  6.61$\pm$0.07 \\
b & 58447.375 &     $K^\prime$ &  11 & 922.6 $\pm$ 2.7 &  68.59$\pm$0.12 &  6.77$\pm$0.06 \\
b & 58447.387 &     $K^\prime$ &   8 & 923.2 $\pm$ 1.7 &  68.60$\pm$0.10 &  6.70$\pm$0.11 \\
b & 58772.590 &     $K^\prime$ &  10 & 908.8 $\pm$ 1.6 &  68.29$\pm$0.10 &  6.56$\pm$0.02 \\
b & 58826.379 &     $K^\prime$ &  10 & 915.6 $\pm$ 1.9 &  68.25$\pm$0.15 &  6.81$\pm$0.07 \\
b & 58826.418 &     $K^\prime$ &  13 & 915.0 $\pm$ 1.8 &  68.52$\pm$0.14 &  6.70$\pm$0.06 \\
b & 59218.387 &     $K^\prime$ &  11 & 911.4 $\pm$ 1.8 &  68.41$\pm$0.10 &  7.21$\pm$0.05 \\
SW & 58352.602 &     $K^\prime$ &   2 & 6246.8 $\pm$ 3.1 & 239.044$\pm$0.013 &  8.55$\pm$0.15$^a$ \\
SW & 58772.590 &     $K^\prime$ &  10 & 6237.4 $\pm$ 1.6 & 239.471$\pm$0.016 &  7.62$\pm$0.05 \\
SW & 59218.387 &      $K$ &  11 & 6220.6 $\pm$ 1.7 & 239.688$\pm$0.015 &  7.71$\pm$0.03 \\
E & 58447.371 &     $K^\prime$ &  15 & 6709.3 $\pm$ 2.4 &  79.027$\pm$0.022 &  6.86$\pm$0.07 \\
E & 58447.387 &     $K^\prime$ &  10 & 6706.8 $\pm$ 1.6 &  79.065$\pm$0.013 &  6.85$\pm$0.05 \\
\hline
\end{tabular}

{\footnotesize $^a$ The isoplanatic patch size was small in this epoch, so it is likely that PSF-fitting photometry did not capture\\
all of the flux from the companion; this value should be disregarded in favor of other measurements.}
\end{table*}

\begin{table}
\begin{center}
\caption{Properties of the 2M0437 Primary Star} 
\label{tab:primary}
\begin{tabular}{l | l | l}
\hline
\multicolumn{3}{c}{Observed Quantities}\\
\hline
$\alpha$ [deg, ICRS] & 69.3405544 & \multirow{8}{0.3\linewidth}{\gaia~EDR3 ID~151499478104075008}\\
$\delta$ [deg, ICRS] & 26.8502960 & \\
$\pi$ [mas] & 7.806 (0.028) & \\
$\mu_{\alpha}$ [mas yr$^{-1}$] & 8.774 (0.035) & \\
$\mu_{\delta}$ [mas yr$^{-1}$] & -27.239 (0.025) & \\
$G$ (Gaia) [mags] & 14.309 (0.003) & \\
$G_{\rm bp}$ (Gaia) & 15.938 (0.006) & \\
$G_{\rm rp}$ (Gaia) & 13.078 (0.004) & \\
\cline{3-3}
$B$ (Johnson) & 17.110 (0.067) & \multirow{2}{*}{APASS DR10}\\
$V$ (Johnson) & 15.661 (0.090) & \\
\cline{3-3}
$g$ (Sloan) & 16.303 (0.003) & \multirow{5}{0.3\linewidth}{Pan-STARRS~DR2 ID~140220693404470986}\\
$r$ (Sloan) & 15.131 (0.006) & \\
$i$ (Sloan) & 13.613 (0.002) & \\
$z$ (Sloan) & 12.915 (0.001) & \\
$y$ (Sloan) & 12.587 (0.003) & \\
\cline{3-3}
$J$ (2MASS) & 11.298 (0.020) & \multirow{3}{0.3\linewidth}{2MASS~PSC J04372171+2651014}\\
$H$ (2MASS) & 10.656 (0.021) & \\
$K_{\rm s}$ (2MASS) & 10.386 (0.018) & \\
\cline{3-3}
$Z$ & 12.328 (0.001) & \multirow{4}{0.3\linewidth}{UKIDSS DR-8 LAS/GCS/DXS J043721.72+265101.2} \\
$Y$ & 11.876 (0.001) & \\
$J$ (MKO) & 11.329 (0.001) & \\
$K$ (MKO) & 10.489 (0.001) & \\
\cline{3-3}
IRAC-I1 (3.5 \micron) & 10.137 (0.002) & \multirow{4}{0.32\linewidth}{\spitzer~Taurus~v2.1 043721.7+265101} \\
IRAC-I2 (4.4 \micron) & 10.074 (0.003) & \\
IRAC-I3 (5.6 \micron) & 10.076 (0.011) &  \\
IRAC-I4 (7.6 \micron) & 9.937 (0.014) &  \\
\cline{3-3}
W1 (3.4 \micron) & 10.284 (0.023) & \multirow{3}{0.3\linewidth}{All-WISE J043721.72+265101.2}\\
W2 (4.6 \micron) & 10.120 (0.021) & \\
W3 (12 \micron) & 10.104 (0.075) & \\
\cline{3-3}
R.V. [km sec$^{-1}$] & 17.1 (0.3) & \multirow{2}{*}{IRD spectra}\\
$v \sin i$ [km s$^{-1}$] & 22.54 (0.14) & \\
\cline{3-3}
$P_{\rm rot}$ [days] & 1.84 & \ktwo\ light curve\\
\halpha [\AA] & 0.54 & SNIFS spectrum\\
\hline
\multicolumn{3}{c}{Inferred Quantities}\\
\hline
$d$ [pc] & 128.1 (0.5) & \multirow{7}{*}{\gaia\ EDR3 + R.V.}\\
$X$ [pc] & -123.62 (0.44) & \\
$Y$ [pc] & 15.57 (0.05) & \\
$Z$ [pc] & -29.79 (0.10) & \\
$U$ [km s$^{-1}$] & +17.0 (0.30) & \\
$V$ [km s$^{-1}$] & -14.05 (0.07) & \\
$W$ [km s$^{-1}$] & -10.43 (0.08) & \\
\cline{3-3} 
\teff\ [K] & 3100 (100) & model fit to spectrum\\
$L_*$ [\lsun]& 0.0686 (0.0005) & model fit to SED\\
$R_*$ [\rsun] & 0.84 (0.11) & Stefan-Boltzmann\\
$[$Fe/H$]$ & +0.01 (0.05) & \cite{DOrazi2011}\\
log~g & 4.5 (0.5) & model fit to spectrum\\
\ebv & $\lesssim$0.1 & SED fit sensitivity\\
$M_*$ [\msun] & $\sim$0.15-0.18 & BHAC15/SPOTS model fits\\
age [Myr] & $\sim$2.5 (0.4) & \cite{Krolikowski2021}\\
\hline
\end{tabular}
\end{center}
\end{table}

\begin{table}
\begin{center}
\caption{Properties of the 2M0437b Companion} 
\label{tab:companions}
\begin{tabular}{l | l | l}
\hline
\multicolumn{3}{c}{Observed Quantities}\\
\hline
$\rho$ [mas] & 918 & \multirow{5}{*}{NIRC2/IRCS AO images + \gaia}\\ 
P.A. [deg] & 68.5 &  \\ 
separation [AU] & 118 (1.3) &   \\
$H$ [mags] & 18.05 (0.08) & \\
$K'$ [mags] & 17.21 (0.07) & \\
\hline
\multicolumn{3}{c}{Inferred Quantities}\\
\hline
$M_p$ [\mjup] & $\sim$4  & \multirow{2}{*}{DUSTY + SONORA models}\\
\teff\ [K] & 1400-1500 & \\
\hline
\end{tabular}
\end{center}
\end{table}

\begin{table}
\begin{center}
\caption{Properties of Co-Moving Star \comover} 
\label{tab:comoving}
\begin{tabular}{l | l | l}
\hline
\multicolumn{3}{c}{Observed Quantities}\\
\hline
$\alpha$ [deg, ICRS] & 69.3597410 & \multirow{8}{0.3\linewidth}{\gaia~EDR3 ID~151593310255495552}\\
$\delta$ [deg, ICRS] & 26.8620293 & \\
$\pi$ [mas] & 8.29 (2.77) & \\
$\mu_{\alpha}$ [mas yr$^{-1}$] & 10.82 (2.23) & \\
$\mu_{\delta}$ [mas yr$^{-1}$] & -27.10 (1.60) & \\
$G$ (Gaia) [mags] & 20.733 (0.014) & \\
$G_{\rm bp}$ (Gaia) & 22.54 (0.24) & \\
$G_{\rm rp}$ (Gaia) & 19.30 (0.05) & \\
\cline{3-3}
$Z$ & 18.562 (0.028) & \multirow{4}{0.3\linewidth}{UKIDSS DR-8 LAS/GCS/DXS J043726.32+265143.4}\\
$Y$ & 17.746 (0.018) & \\
$J$ (MKO) & 16.778 (0.014) & \\
$K$ (MKO) & 15.279 (0.010) & \\
\cline{3-3}
$J$ (2MASS) & 16.639 (0.152) & \multirow{3}{0.3\linewidth}{2MASS~PSC J04372631+2651438}\\
$H$ (2MASS) & 15.620 (0.134) & \\
$K_{\rm s}$ (2MASS) & 15.144 (0.145) & \\
\cline{3-3}
IRAC-I1 (3.5 \micron) & 14.91 (0.01) & \multirow{4}{0.32\linewidth}{\spitzer~SSTSL2 J043726.31+265143.5} \\
IRAC-I2 (4.4 \micron) & 14.83 (0.02) & \\
IRAC-I3 (5.6 \micron) & 14.44 (0.06) &  \\
IRAC-I4 (7.6 \micron) & 14.53 (0.18) &  \\
\cline{3-3}
W1 (3.4 \micron) & 15.016 (0.036) & \multirow{3}{0.3\linewidth}{All-WISE J043726.33+265143.3}\\
W2 (4.6 \micron) & 14.936 (0.089) & \\
\hline
\end{tabular}
\end{center}
\end{table}

\begin{table*}
\caption{Detection Limits for Additional Companions to 2M0437 \label{tab:ImgLim}}
\begin{tabular}{lrcrrrrrrrrrrrr}
\hline
Name & Epoch & Filter & $N_{obs}$ & $t_{int}$ & \multicolumn{10}{c}{Contrast $\Delta m$ (mag) at $\rho = $ (mas)} \\
& (MJD) & & & (sec) & 150 & 200 & 250 & 300 & 400 & 500 & 700 & 1000 & 1500 & 2000 \\
\hline
      EPIC 248131102 & 58352.60 &   Kp     & 12 &  240.00 &  5.3 &  6.0 &  6.2 &  6.9 &  7.4 &  8.0 &  8.5 &  9.3 &  9.4 &  9.4 \\
    EPIC 248131102 & 58826.41 &   Kp     & 17 &  970.00 &  3.8 &  4.9 &  5.1 &  5.6 &  6.1 &  6.6 &  7.9 &  9.8 & 10.5 & 10.6 \\ 
    \hline
\end{tabular}
\end{table*}

\section*{Acknowledgements}

We thank H. Nayyeri, A. S. Long, and A. R. Cooray for obtaining a NIRC2 observation, and A. Fukui for assistance with IRCS data reduction.  E. G. acknowledges support from NASA Exoplanets Research Program award 80NSSC20K0957.  T.A.B. acknowledges support from a NASA FINESST award (80NSSC19K1424).  
A.L.K. acknowledges support from NASA Keck PI Data Awards, administered by the NASA Exoplanet Science Institute.   Some data presented herein were obtained at the W. M. Keck Observatory, which is operated as a scientific partnership among the California Institute of Technology, the University of California and the National Aeronautics and Space Administration. The Observatory was made possible by the generous financial support of the W. M. Keck Foundation. This research made use of Astropy,\footnote{http://www.astropy.org} a community-developed core Python package for Astronomy \citep{Astropy2013,astropy:2018}. This research has made use of the NASA/IPAC Infrared Science Archive, which is funded by the National Aeronautics and Space Administration and operated by the California Institute of Technology.  This publication makes use of VOSA, developed under the Spanish Virtual Observatory project supported by the Spanish MINECO through grant AyA2017-84089.  VOSA has been partially updated by using funding from the European Union's Horizon 2020 Research and Innovation Programme, under Grant Agreement number 776403 (EXOPLANETS-A). This work is partly supported by JSPS KAKENHI Grant Numbers JP19K14783, JP21H00035, JP18H05442, JP15H02063, and JP22000005. The data analysis was carried out, in part, on the Multi-wavelength Data Analysis System operated by the Astronomy Data Center (ADC), National Astronomical Observatory of Japan.  {\bf Data Availability:}  All ground-based data obtained for this project are available from the authors (UH88/SNIFS) or the Subaru SMOKA, Keck KOA, or IRTF archives.  \ktwo\ data is available for download from the MAST Archive at STScI.




$\bibliographystyle{mnras}
$\bibliography{references} 

\bsp	
\label{lastpage}
\end{document}